\documentclass{jfm}
\usepackage{graphicx}
\usepackage{epstopdf, epsfig}
\usepackage{caption}
\usepackage{subcaption}
\usepackage{amssymb}
\usepackage{amsmath}
\usepackage{scalerel}
\usepackage{xcolor}
\usepackage{cancel}
\usepackage{comment}

\shorttitle{Unsteady turbulent separation over a 3D Gaussian bump}
\shortauthor{K.H. Manohar, H. Annamalai, O. Williams, C. Morton, and R.J. Martinuzzi}

\title{Unsteadiness in turbulent separated flow over a three-dimensional Gaussian bump}

\author{Kevin H. Manohar\aff{1}
  \corresp{\email{kmanoh@uw.edu}},
  Hariprasad Annamalai\aff{1},
  Owen Williams\aff{1},
  Chris Morton\aff{3},
  \and Robert J. Martinuzzi\aff{2}}

\affiliation{\aff{1}William E. Boeing Department of Aeronautics and Astronautics, University of
Washington, Seattle, WA 98195, U.S.A.
\aff{2}Department of Mechanical and Manufacturing Engineering, University of Calgary, Calgary, AB T2L 1Y6, Canada
\aff{3}Department of Mechanical Engineering, McMaster University, Hamilton, ON L8S 3L8, Canada 
}

\begin{document}

\maketitle

\begin{abstract}
The unsteady separated flow over the three-dimensional \textit{Boeing Gaussian Bump} is investigated at a Reynolds number based on bump height $\Rey_H = 2.26\times10^5$ using unsteady wall-pressure measurements and planar particle image velocimetry (PIV). 
Four major unsteady broadband phenomena spanning more than two decades in frequency are identified:
\textbf{(1)} a very-low-frequency (VLF) spanwise  motion centered at a Strouhal number of $St_H\sim10^{-3}$ (1~Hz) based on bump height,
\textbf{(2)} a low-frequency breathing motion of the separation zone centered at $St_{L_{\rm{sep}}}=0.068$ (13.5~Hz) where $L_{\rm{sep}}$ is the mean separation length,
\textbf{(3)} a 20~Hz frequency that appears to be associated with vortex shedding from the lateral shear layers, and
\textbf{(4)} a centreline shear-layer vortex shedding at $St_{L_{\rm{sep}}}=0.68-1.01$ (135-200~Hz). 
Interestingly, while the VLF mode has a characteristic frequency of the same order to that often reported for other rectilinear bodies and hills that exhibit bistable asymmetric wake-switching, it is found that the VLF mode for this geometry exhibits a continuous spanwise meandering motion. 
Joint symmetric-antisymmetric proper orthogonal decomposition modal statistics from top-down PIV data further show that the spanwise meandering and streamwise stretching of the wake---likely associated with the breathing motion---are dynamically coupled, with the separation zone reaching its greatest streamwise extent when in a symmetric state.
In this paper, the observed hierarchy of spectral features is comparable with those observed for a wide range of geometries, suggesting connections between geometric lengthscales and the low-frequency dynamics.

\end{abstract}

\section{Introduction\label{sec:Intro}}
The presence of adverse pressure gradients (APGs) \citep{sears1975} and surface curvature \citep{bradshaw1973} around streamlined and bluff bodies can give rise to three-dimensional (3D) separation and associated unsteady instabilities that span multiple frequency bands. The resulting separated flow is highly complex due to the coupling between the unsteady dynamics of the recirculating region, large-scale coherent motions, incoming turbulence, and the competing influences of pressure gradients and curvature.
Although these effects jointly determine both the mean and unsteady behavior, our current physical understanding of 3D turbulent separation remains incomplete---particularly for flows in which the separation front is not fixed but free to adjust to the local pressure field and curvature. In the present work, we examine the unsteady 3D dynamics of turbulent separation over the smooth \textit{Boeing Gaussian Bump} with error function shoulders, with an emphasis on explorations of the low-frequency motions.

\vspace{0.5em}
\noindent
Bumps and hills subjected to turbulent boundary-layer inflow \citep[see, e.g.,][]{williams2022} provide canonical test cases to investigate the combined influence of pressure gradient and surface curvature on smooth-body flow separation, which together produce far more complex dynamics than either in isolation. This challenge has motivated the development of a new generation of benchmark configurations, such as the \textit{Gaussian Bump} \citep{slotnick2019,williams2020} and the BeVERLI hill \citep{gargiulo2021}, among others, established to improve the modeling and prediction of turbulent separation as part of NASA’s CFD Vision 2030 initiative \citep{vision2030}. 
The \textit{Gaussian Bump} features a smooth Gaussian profile with a large spanwise aspect ratio (width-to-height, $B_w/H\approx8.3$), producing a wide separated footprint with pronounced 3D features \citep{williams2020, williams2021}. The flow behavior over such geometries is sensitive to shape parameters, particularly the aspect ratio ($B_w/H$) and the relative height of the bump compared with the incoming boundary-layer thickness, which together influence the reorientation of vorticity \citep{mason1987}, the distribution of upwash and downwash in the wake, and the proportion of mass flux that passes over versus around the body. Most of what is currently known about hill-type turbulent separated flows concerns the mean topology; comparatively little is known about the organization of the unsteady motions that give rise to that mean.

\vspace{0.5em}
\noindent
Although the mean topology of hill-type separated flows has been extensively characterized \citep[e.g.][]{garcia2009,williams2021,gargiulo2021,duetsch2022,simmons2022,simmons2024}, it is the underlying unsteady dynamics that govern the body-force fluctuations, organize the Reynolds stresses, and mediate the inter-scale energy transfers that ultimately sustain the mean flow structure.
Even in canonical 2D wakes such as the cylinder at moderate Reynolds numbers \(\Rey>300\), the mean flow does not exist instantaneously—the familiar recirculation zone arises from the time-averaged imprint of self-sustained vortex shedding \citep{williamson1996}. In 3D smooth-body separations, the situation is considerably more intricate: the separation front and reattachment region undergo broadband motions involving both shear-layer instabilities and large-scale low-frequency unsteadiness \citep{eaton1982,byun2006,weiss2015,ching2020}. These unsteady dynamics, rather than the mean field alone, determine the exchange of momentum between the separated shear layer, the recirculation zone, and the outer boundary layer \citep{kiya1983,driver1987,wu2020}. However, most previous studies of hill and bump flows have focused on time-averaged quantities, leaving the nature and spectral organization of the unsteady motions—particularly those at low frequencies—much less understood.

\vspace{0.5em}
\noindent
At high Reynolds numbers, separated turbulent flows exhibit a broad hierarchy of time scales \citep{eaton1982,hudy2003,weiss2015}. In nominally two-dimensional configurations, two dominant modes are well established. The first is a medium-frequency \emph{vortex-shedding} mode associated with the roll-up and advection of spanwise vortices \citep{na1998,weiss2015}, of order $St_{L_{\rm{sep}}}=fL_{\rm{sep}}/U_{\infty}=O(10^{-1})$, with $L_{\rm{sep}}$ the mean separation length. The second is a lower-frequency \emph{flapping} or \textit{breathing} mode, corresponding to an oscillatory vertical motion of the separated shear layer \citep{eaton1982,cherry1984,driver1987,weiss2015,wang2022}, typically an order of magnitude slower than the shedding frequency ($St_{L_{\rm{sep}}}=O(10^{-2})$). This low-frequency motion has been described either as \textit{flapping}---a vertical oscillation of the separated shear layer in geometry-induced turbulent separation bubbles (TSBs) that modulates the reattachment point---or as \textit{breathing}---a modulation of the size of the recirculation zone in variable-separation TSBs where both separation and reattachment points fluctuate. Although closely related, we use the term \textit{breathing} in the context of variable-separation flows here to emphasize changes in the separated zone extent rather than a purely kinematic shear-layer deflection. A growing body of work also suggests that, even in nominally 2D configurations, such low-frequency motions may be manifestations of an underlying 3D global spanwise stationary mode \citep{borgmann2024, steinfurth2025, fuchs2025}.

\vspace{0.5em}
\noindent
In 3D geometries, additional long time-scale dynamics emerge. In particular, a \emph{very-low-frequency} (VLF) mode associated with instantaneous asymmetric motions of the wake at $St=O(10^{-3})$ has been reported in axisymmetric wakes \citep{rigas2014,pavia2019} and in rectilinear-type bodies such as the Ahmed body, a surface-mounted half-axisymmetric body and the BeVERLI hill \citep{grandemange2013,pavia2018,duetsch2022,panesar2023}. This mode is typically one order of magnitude slower than the breathing mode and is associated with a global spanwise instability—either as an azimuthal precession in axisymmetric bodies or as spanwise-bistable switching between mirror-symmetric states in rectilinear ones \citep{rigas2014,pavia2018,grandemange2012a,grandemange2013,macgregor2023}. The stability of these motions is highly sensitive to geometry: small variations in aspect ratio or base curvature can trigger reflectional-symmetry breaking \citep{garcia2009,grandemange2013,duetsch2022,panesar2023}.

\vspace{0.5em}
\noindent
The present study explores the properties of low/mid-frequency and three-dimensional unsteady motions on the \textit{Gaussian Bump} using a combination of simultaneously sampled unsteady surface pressures and particle image velocimetry (PIV) in orthogonal planes. \S\ref{sec:Intro_unsteadiness} reviews prior observations of unsteady modes in two- and three-dimensional separated flows. The geometry and past work on the bump are briefly reviewed in \S\ref{sec:Intro_bump}. \S\ref{sec:Methods} introduces the measurement techniques, experimental setup, and the symmetrization method used to extract symmetric and antisymmetric velocity components. The mean field is presented in \S\ref{sec:RD_mean}, followed by an analysis of the unsteady dynamics in \S\ref{sec:RD_unsteady}. A broader discussion of the three-dimensional unsteady dynamics is provided in \S\ref{sec:discussion}, and the key findings are summarized in \S\ref{sec:Conclusions}.


\section{Background: Unsteadiness in turbulent separated flows\label{sec:Intro_unsteadiness}}
A brief review of the relevant literature on the key unsteady modes found in nominally 2D flows is provided in \S\ref{sec:2Dintro}, and in 3D flows in \S\ref{sec:3Dintro}.

\subsection{Unsteadiness in two-dimensional flows\label{sec:2Dintro}}
The unsteadiness of turbulent separated flows developing from geometrical singularities—such as backward-facing steps, blunt plates, and fence-and-splitter configurations—has been well documented \citep[e.g.][]{kiya1983,eaton1982,cherry1984,driver1987,spazzini2001}. In these flows, the mean streamlines form a recirculation region that reattaches to the wall, defining a TSB \citep{eaton1982,kiya1983}. This closed bubble topology, however, exists only in the statistical mean: instantaneous realizations exhibit non-reattaching topologies, with continuous mass exchange between the separated region and the outer stream \citep{kiya1983,weiss2015,wu2020}. The intermittent opening and closing of this region is mediated by spanwise instabilities of the shear layer, which give rise to low-frequency flapping motions observed in fixed-separation flows, breathing motions in variable-separation flows, and to medium-frequency vortex shedding modes \citep{eaton1982,cherry1984,driver1987,weiss2015,wang2022,fang2024}. Across a range of geometry-induced TSBs, the corresponding Strouhal numbers typically fall in the ranges $St_{L_{\rm{sep}}}=0.08$–$0.2$ for the flapping mode \citep{fang2024} and $0.5$–$1.0$ for the shedding mode \citep{eaton1982, kiya1983, cherry1984, na1998dns, hudy2003, weiss2015,wu2020}. The breathing motion has also been reported in pressure-induced and variable-separation TSBs, with $St_{L_{\rm{sep}}}\approx0.01$ \citep{weiss2015,taifour2016}, and $St_{L_{\rm{sep}}}\approx0.05$ in trailing-edge separations \citep{wang2022}. 

\vspace{0.5em}
\noindent
The flapping mode has been extensively reported in geometry-induced, fixed-separated flows \citep[e.g.][]{eaton1982,kiya1983,cherry1984,driver1987}. It appears to manifest as vertical oscillations of the separated shear layer, leading to streamwise oscillations of the reattachment point. 
In adverse-pressure-gradient separations, the freely moving separation front introduces an additional degree of freedom, yielding broadband low-frequency oscillations associated with the TSB expansion and contraction, often termed \textit{breathing}. 
Several mechanisms have been proposed to identify potential sources of the flapping \citep{eaton1982,kiya1983,cherry1984} and breathing modes \citep{wu2020,wang2022,fang2024}, and is subject of ongoing research.
Moreover, it appears that neither the flapping, breathing nor shedding frequencies collapse universally with $U_{\infty}/L_{\text{sep}}$, since $L_{\text{sep}}$ itself evolves with the unsteady state of the bubble and local shear-layer development \citep{wu2020}.

\vspace{0.5em}
\noindent
Recent studies on smooth-geometry, nominally 2D TSBs further show that the low-frequency flapping and breathing motions observed in a symmetry plane in fixed- and variable-separation TSBs, respectively, is often the planar manifestation of an inherently 3D stationary global mode \citep{cura2024,borgmann2024,steinfurth2025,cura2025,fuchs2025}. 
For wall-mounted humps, \citet{borgmann2024} find that the instantaneous separation location varies only weakly along the surface. In particular, despite being a variable-separation flow, they find that the dominant low-frequency shear-layer dynamics ($St_{L_{\rm{sep}}}\lesssim0.1$) resemble the flapping mode reported for geometry-induced TSBs, and they are in fact the 2D projection of a spanwise meandering structure. Using time-resolved PIV and spectral POD, they show that these spanwise-undulating structures possess wavelengths in the range $0.62\,L_{\rm{sep}} \lesssim \lambda_y \lesssim 1.26\,L_{\rm{sep}}$ (i.e.\ $\lambda_y = \mathcal{O}(L_{\rm sep})$), extend across the full streamwise length of the TSB, and generate alternating peaks and valleys in the recirculating region. The spatial structure of these 3D modes---and their associated frequency ($St_{L_{\rm{sep}}} \approx 0.1$, near the lower bound of the geometry-induced TSB flapping range $St_{L_{\rm{sep}}} \approx 0.08$–$0.20$)---is consistent with the dominant eigenfunctions associated with the shear-layer flapping mode observed in the centreline-symmetry plane.
Similarly, time-resolved measurements over a smooth backward-facing ramp by \citet{steinfurth2025} reveal that the low-frequency band ($St_{L_{\rm{sep}}}\lesssim0.05$) corresponds to a 
spanwise-structured meandering mode: a mirror-inverted pattern that is locked in space while oscillating in time, with spanwise wavelength on the order of the channel width, $\lambda_y=O(L)$, and pronounced nodes and antinodes set by the tunnel sidewall confinement.
For a straight backward-facing ramp, \citet{fuchs2025} used resolvent analysis to show that such stationary spanwise modes can arise from the interference of left- and right-travelling 3D modes reflected by the tunnel sidewalls, with dominant spanwise wavelengths of several TSB lengths.
In both ramp configurations, the streamwise-vertical expansion and contraction observed in the symmetry plane is interpreted as the projection of these spanwise-structured modes whose wavelengths are governed primarily by the spanwise domain size $O(L)$ (which is typically several times $L_{\rm{sep}}$).
These recent results support the view that the apparently 2D breathing motion seen in smooth-body separations is often the planar imprint of an underlying 3D stationary global mode, and that the spanwise domain width, $L$, emerges as a key length scale alongside the mean separation length, $L_{\rm{sep}}$, when interpreting low-frequency unsteadiness.

\subsection{Three-dimensional unsteadiness \label{sec:3Dintro}}
In addition to the shedding and breathing motions characteristic of two-dimensional separated flows, three-dimensional (3D) wakes have sometimes been observed to exhibit an additional \textit{very-low-frequency (VLF)} mode associated with large-scale spanwise or azimuthal unsteady motions of the wake. This motion occurs on time scales one to two orders of magnitude slower than the dominant shedding or breathing motions and represents a defining feature of 3D turbulent separation. Its form appears to depend strongly on geometry. 

\vspace{0.5em}
\noindent
Axisymmetric bodies exhibit both axial {bubble-pumping motions} and a slow azimuthal precession of the wake \citep{rigas2014,pavia2019,zhang2023,panesar2023}. 
For such geometries, the wake may freely precess around the body, admitting an infinite continuum of equivalent asymmetric states corresponding to random rotations of the planar vortex-shedding axis \citep{rigas2014}—i.e.\ a stochastic meandering along the azimuthal coordinate. 
The VLF motion originates from a symmetry-breaking bifurcation of the steady axisymmetric wake \citep{fabre2008}, yielding a planar-symmetric state that subsequently undergoes stochastic precession at \(St_D \sim 10^{-3}\) (with body diameter $D$)—an order of magnitude slower than the bubble-pumping motion (\(St_D \sim 10^{-2}\)) and two orders slower than the vortex-shedding mode (\(St_D \sim 10^{-1}\)) \citep{pavia2019,zhang2023}.

\vspace{0.5em}
\noindent
Rectilinear geometries, such as the Ahmed body and the BeVERLI hill, restrict symmetry-breaking modes to reflection about a plane, producing a bistable wake, where the wake switches intermittently between two spanwise-asymmetric states \citep{grandemange2012a,grandemange2013,duetsch2022,macgregor2023}. 
In such flows, the wake typically dwells for extended periods in one asymmetric configuration before randomly switching to the other.

Despite the different symmetry constraints, both axisymmetric and rectilinear wakes exhibit comparable VLF frequencies (\(St_D \sim 10^{-3}\)).
In contrast, the intermediate bubble-pumping motion (\(St_D \sim 10^{-2}\))—well established in axisymmetric and surface-mounted bluff bodies \citep{pavia2018,panesar2023}—has not previously been observed in smooth hill- or bump-type configurations.

\vspace{0.5em}
The discussion above motivates the following key questions that will be addressed in the present study:
\begin{enumerate}
    \item What are the dominant unsteady modes in the \textit{Gaussian Bump} flow? Specifically, does the flow exhibit the VLF spanwise, breathing, and vortex-shedding modes previously identified in three-dimensional turbulent separations, and how do their spatial and spectral characteristics compare across these frequency bands?
    \item If observed, what is the nature of the VLF dynamics? In particular, are these motions associated with spanwise fluctuations of the wake, and do they resemble the continuous precessional (meandering) behavior of axisymmetric wakes or the discrete bistable switching seen in rectilinear bodies?
\end{enumerate}

\section{The Boeing Gaussian Bump\label{sec:Intro_bump}}
\begin{figure}
    \centering
    \includegraphics[width=0.7\textwidth]{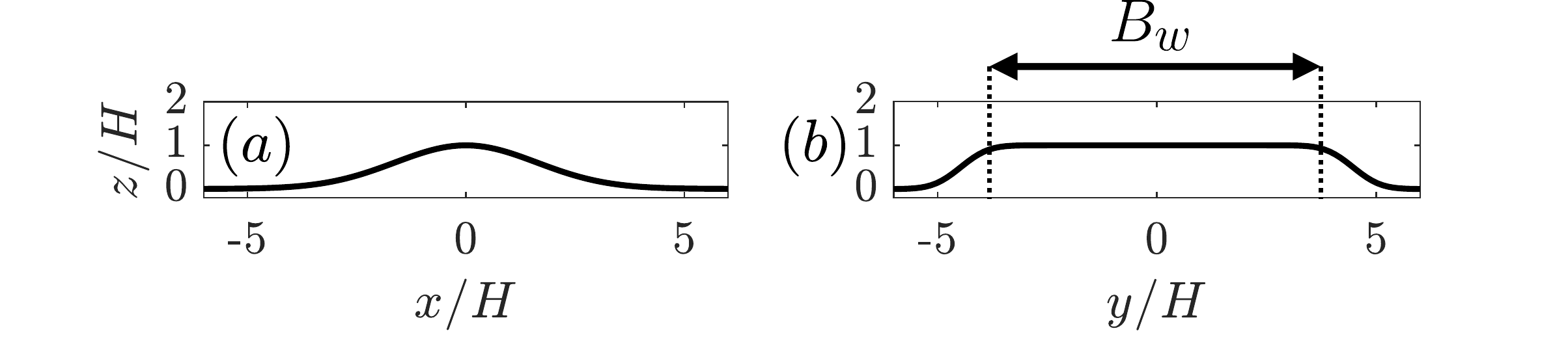}
    \caption{\small{2D profiles of the Boeing Gaussian Bump along the two symmetry planes: $(a)$ $y/H=0$ (flow from left to right); $(b)$ $x/H = 0$, with annotation of Bump width $B_w$.}}
    \label{fig:Bump2Dprofiles}
\end{figure}

The surface of the \textit{Boeing Gaussian Bump} is defined as
\begin{equation}\label{eqn:bump_profile}
    z(x,y) = \frac{H}{2} \, e^{(-x/x_0)^2}
    \Big[ 1 + \mathrm{erf}\!\left( \frac{L_b/2 - 2y_0 - |y|}{y_0} \right) \Big],
\end{equation}
where \(x\), \(y\), and \(z\) denote the streamwise, spanwise, and vertical directions, respectively.
The splitter-plate width is \(L_b = 35.5'' = 0.9017~\mathrm{m}\), defining the geometric parameters \(x_0 = 0.195\,L_b\), \(y_0 = 0.06\,L_b\), and bump height \(H = 0.085\,L_b = 0.0766~\mathrm{m}\).
The full-width, half-maximum bump width is \(B_w = 8.33H = 0.640~\mathrm{m}\).
Profiles along the two symmetry planes, \(y/H = 0\) and \(x/H = 0\), are shown in figure~\ref{fig:Bump2Dprofiles}.
The wind-tunnel ceiling was located at half the tunnel width above the inflow splitter plate, establishing the intended confinement.

\vspace{0.5em}
\noindent
The geometry is designed to replicate challenges relevant to high-lift wing configurations ($H/\delta\sim O(10)$) and the spanwise taper aims to limit complex side-wall interactions. \cite{williams2020} performed oil-film surface flow visualizations on the Bump using china clay and kerosene in the $3'\times3'$ Low-Speed Wind Tunnel at the University of Washington, revealing the 3D nature of the flow. Surface-pressure measurements revealed that the flow is Reynolds-number insensitive in the mean for $\Rey_H\,\geq\,226\,000$. The flow is dominated by two counter-rotating surface vortices on either side of the centreline symmetry plane $y/H=0$. The shear-stress surface pattern resembles the ubiquitous owl-face-of-the-first-kind pattern, consisting of a saddle separation on the centerline that splits into two symmetric focii (nodes) on either side, and another saddle point on the centerline downstream. The two counter-rotating vortices emerge approximately at the location of surface-concavity change ($x/H\approx1.67$). This pattern was also reproduced by \cite{gray2021} at a higher Reynolds number ($\Rey_H=335\,000$).

\vspace{0.5em}
\noindent
The off-surface vortex topology of the owl-face-of-the-first-kind was conceptualized by \cite{perryHornung1984}---consisting of two vorticies emanating from the focii (both of opposite sense) being tilted into forming a streamwise-oriented counter-rotating vortex pair. Recent stereoscopic PIV work by \cite{gray2023} shows agreement with this conceptual picture.

\vspace{0.5em}
\noindent
Comparison of experimental and computational results indicate that the Bump presents a significant challenge for predicting flow separation, with Reynolds-Averaged Navier-Stokes (RANS) simulations and Delayed Detached Eddy Simulations (DDES) producing a much thinner separation zone, opposite Reynolds number trends and sometimes asymmetric results \citep{williams2020,gray2021,gray2022}. 

\vspace{0.5em}
\noindent
Recently, \cite{manohar2023} developed a machine-learning-based technique to temporally super-resolve under-sampled PIV snapshots of the centreline Bump flow ($\Rey_{H}=226\,000$) around separation. Their work facilitated the interpretation of motions governing the multiple frequencies measured by the wall-pressure sensors. They found that a vortex shedding frequency $St_{L_{\rm{sep}}}=0.67$, is present along the centreline plane, and is potentially associated with the formation, agglomeration and advection of large-scale vortical structures. A lower frequency at $St_{L_{\rm{sep}}}=0.067$, present upstream of the separation front and just downstream of the apex at the centreline symmetry plane, was interpreted to reflect the \textit{breathing} vertical oscillation of the separated shear layer.  

\vspace{0.5em}
\noindent
The present work extends these investigations to consider dynamics at the very lowest frequencies observed in the wall-pressure spectra.
The analysis characterizes the VLF component, examines its relationship to the higher-frequency breathing and shedding motions, and evaluates its connection to the bi-stable or meandering behavior reported in other hill and bluff-body wakes.
By combining synchronized wall-pressure and PIV measurements, the work aims to provide a unified description of the multi-frequency dynamics that govern the unsteady separation over the \textit{Gaussian Bump}.

\section{Methods\label{sec:Methods}}
Experiments were conducted in the \(3'\times3'\) Low-Speed Wind Tunnel at the University of Washington. A rendering of the \textit{Gaussian Bump} model and the planar PIV configurations in two orthogonal orientations is shown in figure~\ref{fig:PIV_setups}. The splitter plate spans \(L_b = 0.9017~\mathrm{m} = 11.8\,H\) in width, with adhesive-backed foam strips sealing the gap to the tunnel side walls so that the bump spans the full tunnel width $L=0.9144~\mathrm{m} = 11.94\,H$. The 3D-printed leading-edge section, of length \(8'' = 0.203~\mathrm{m} = 2.65\,H\), follows a 10:1 modified super-ellipse profile and is positioned such that the bump apex (\(x=0\)) lies one plate width \(L\) downstream of the stagnation point. A 240-grit sandpaper strip located at \(x = -0.778\,L = -9.28\,H\) serves as a tripping device to ensure a fully turbulent boundary layer. 
The stagnation point was set just above the geometric leading edge using a \(12'' = 0.305~\mathrm{m} = 3.98\,H\) trailing-edge flap deflected by \(7^{\circ}\), whose angle was set by comparing pressure taps symmetrically placed either side of the leading edge. The freestream velocity, \(U_{\infty}\), was monitored using an Omega PX653-10D5V differential pressure transducer measuring the static pressure difference between the inlet and test-section entrance. The transducer was calibrated using a pitot static tube above the plate at the BL measurement location. BL mean velocity profiles were obtained at \(x = -0.43\,L = -5.13\,H\) using a pitot probe on a traversing mechanism \citep{williams2020}, and the corresponding parameters are listed in table~\ref{tab:BL}. Each dataset comprises synchronized high-rate wall-pressure and planar PIV measurements, acquired at multiple orthogonal planes as described in the following subsections. The inflow Reynolds number was chosen as surface-pressure measurements were found to be $\Rey$-invariant in the mean for $\Rey_H \geq 226\,000$ \citep{williams2020}.

\begin{table}
  \centering
  \begin{tabular}{cccccc}
     $U_{\infty}$ [$\text{ms}^{-1}$] &  $H/\delta$ & $H/\theta$ & $\Rey_{\theta}$ & $\Rey_H$ & $Re_L$\\
    \hline
     45  & 12.7 & 106 & 2100  & 226\,000 & $2.70 \times 10^{6}$  \\
  \end{tabular}
  \caption{\small Inflow and BL characteristics; $U_{\infty}$ the freestream speed measured, $\delta$ is the 99\% thickness of the incoming boundary layer and $\theta$ its momentum thickness. Measurement details are provided in \cite{williams2020}. The nominal Reynolds number based on $U_{\infty}$ and the Bump height, $H\,=\,0.0766$ m, is $\Rey_H\,=\,2.26\times10^{5}$.
}
  \label{tab:BL}
\end{table}

\subsection{Instrumentation setup\label{sec:Methods_pressureSetup}}
Static wall pressure was measured at 25 surface locations (figure~\ref{fig:Cp}) using AllSensors D1-P4V miniature amplified differential pressure transducers with full-scale ranges of 5" and 10"~$\mathrm{H_2O}$. The transducers were sampled through a multiplexed Advantech PCIE-1805 data acquisition card. Each sensor has a manufacturer-specified flat frequency response up to 10~kHz.

\vspace{0.5em}
\noindent
All pressure taps were drilled normal to the surface with a diameter of \(1/32'' = 0.79~\mathrm{mm}\), onto which 2" (\(51~\mathrm{mm} = 0.67\,H\)) long stainless-steel tubes of 1/16" outer diameter were epoxied. Each tube was connected to the measurement port of a transducer via approximately 2" (\(0.67\,H\)) of 1/16" inner-diameter polyurethane tubing. The reference ports of all transducers were routed through 20" (\(508~\mathrm{mm} = 6.63\,H\)) polyurethane tubing to a common manifold connected to an upstream centreline static tap located at \(x/L = -0.8299\) (Tap~1 in \cite{williams2020}).

\vspace{0.5em}
\noindent
The wall-pressure signals were sampled at 31.25~kHz with a timing accuracy of 0.032$~\mathrm{ms}$ \((1/31.25)\). The total length of acquisition is 4.44 min, enabling sufficient resolution of the very-low-frequencies. 
Signals were low-pass filtered using a minimum-order digital filter with a cutoff frequency of 1~kHz. 
Premultiplied wall-pressure spectra (figure~\ref{fig:CpRMS_spectra}) were computed from \(1.72\times10^6\) samples acquired at 31.25~kHz using Hamming windowing with 50\% overlap. Window sizes of \(2^{18}\) and \(2^{11}\) samples were used for very-low-frequency and high-frequency analyses, respectively, to balance resolution and spectral smoothing.

\vspace{0.5em}
\noindent
The wind tunnel is powered by a constant-RPM, 16-blade variable-pitch fan, which introduces two spurious frequencies in all spectra: 
(i) a sharp 476~Hz peak corresponding to the blade-passing frequency (\(f_{\mathrm{RPM}}/16\)), and 
(ii) a broadband subharmonic near 30~Hz (476 Hz/16). 
Both features are most pronounced in taps upstream of the bump apex. A narrow notch filter at 476~Hz was applied to all wall-pressure signals to remove the blade-passing tone.

\subsection{Particle Image Velocimetry setup \& image processing\label{sec:Methods_PIVsetup}}

\begin{figure}
    \centering
    \includegraphics[width=\textwidth]{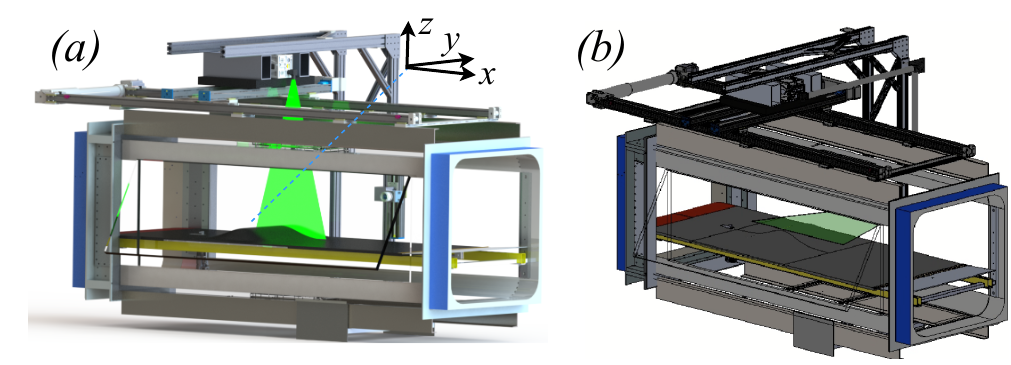}
    \caption{\small{$(a)$ Rendering of streamwise-vertical ($x$-$z$) PIV setup with coordinate system, Bump model and splitter plate in $3'\times3'$ test section [Adapted from \cite{robbins2021}]; $(b)$ Streamwise-spanwise ``top-down" ($x$-$y$) PIV setup [Reproduced from \cite{annamalai2022}].}}
    \label{fig:PIV_setups}
\end{figure}

\begin{figure}
    \centering
    \includegraphics[width=0.8\textwidth]{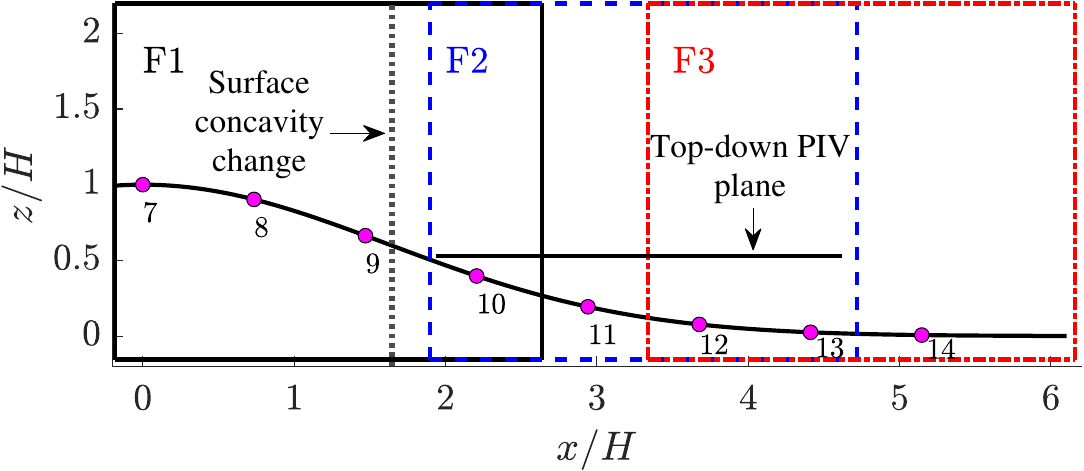}
    \caption{\small{Location and size of the three overlapping FOVs - F1, F2, and F3 ($\geq 20\%$ overlap) - along the streamwise-vertical $x$-$z$ plane that is used for all spanwise measurement planes $y/H=0,\,0.75,\,1.47$. Magenta dots show location of centreline ($y/H=0$) wall-pressure taps labelled 7-14. Streamwise location of surface concavity change at $x/H=1.65$ noted as well as position and streamwise extent of ``top-down" $x$-$y$ PIV plane at $z/H=0.53$.}}
    \label{fig:xz_FOVs}
\end{figure}

A series of planar, two-component, single camera PIV experiments are conducted in two orthogonal orientations: three streamwise-vertical (``side-on view", $x$-$z$) planes $y/H\,=\,0,\,0.75,\,1.47$ and a single streamwise-spanwise plane (``top-down" view, $x$-$y$) at $z/H\,=\,0.53$, focusing on the separated region just downstream of the surface vortices. A rendering of the two setups is provided in figure \ref{fig:PIV_setups} and the locations of the PIV frames are shown in figure \ref{fig:xz_FOVs}. The camera and laser were mounted to a two-axis traversing system, allowing for the adjustment of measurement locations without the need for calibration. The side-on velocity fields are measured independently and with 20\% overlap allowing the creation of combined statistical fields during post-processing (see figure \ref{fig:xz_FOVs}). Only a single frame was acquired, overlapping with the model centerline but biased toward the positive \textit{y}-direction. For both configurations, images were acquired with an Imager sCMOS camera (2560pix x 2160pix) equipped with a Nikon Micro 60 mm focal length lens with an aperture set to f/8. This resulted in fields-of-view (FOVs) of  $22.25 \times 18.77$ $\rm{cm}^2$ (11.14 pixels/mm)  and $20.5 \times 24.5$ $\rm{cm}^2$ (10.46 pixels/mm) in the side-on and top-down orientations, respectively.

\vspace{0.5em}
\noindent
Illumination was provided by a Quantel Evergreen dual-pulsed 200 mJ $\rm{pulse}^{-1}$ 532 nm Nd: YAG laser with a maximum repetition rate of 15 Hz, producing a $\approx1$ mm thick light sheet. A total of $10\,000$ snapshots were acquired across three independent runs for the three side-on FOVs, whereas $4\,000$ snapshots are used for the top-down dataset. The length of these datasets was found to be sufficient to achieve statistical convergence in the first- and second-order moments. The flow was seeded with a mixture of de-ionized water and glycol smoke particles, with a nominal size of 1 $\mu$m.

\vspace{0.5em}
\noindent
The entire image acquisition process, calibration, and post-processing using PIV were controlled through the LaVision DaVis 10 software. The time interval between consecutive laser pulses was adjusted to limit the maximum particle displacement to 12 pixels, resulting in inter-frame times of 25.4 $\mu$s and 28.63 $\mu$s for the side-on and top-down FOVs, respectively. 

\vspace{0.5em}
\noindent
Images were processed using a multi-grid, multi-pass, cross-correlation approach. The final pass utilizes a 32 $\times$ 32 pixel interrogation window with Gaussian windowing and 75\% overlap. The spatial resolution is estimated to be 26.5 pixels per interrogation window, equivalent to 2.38 mm ($=0.03H$) (see \citet{williams2014}). The resulting vectors are validated using a correlation level filter (r $<3$) and three passes of a normalized median filter with a 5 $\times$ 5 vector window, following the approach by \cite{westerweel2005}. The vector validation procedure rejected $1.66\% \pm 0.32\% $ of vectors per frame in the top-down dataset and $0.44\% \pm 0.11\%$ per frame in the side-on dataset.

\vspace{0.5em}
\noindent
Random PIV measurement noise was estimated using a classical near-zero spatial autocorrelation intercept approach, which was demonstrated for PIV by \cite{discetti2013}. For each velocity component, the normalized spatial autocorrelation $R(\Delta)$ was computed and averaged within local regions of the flow-field, and a symmetric parabola was fitted to $R(\Delta)$ at small, non-zero separations. Extrapolating this fit to $\Delta=0$ provides an estimate of the noise variance fraction because turbulent fluctuations remain spatially correlated at small separations, whereas measurement noise contributes to the variance but negligibly to the covariance at non-zero separation. Consequently, the measured correlation falls below unity near the origin.
If the total streamwise velocity fluctuation variance is $\overline{u'^2}=\sigma_s^2+\sigma_n^2$ where $\sigma_s^2,\,\sigma_n^2$  are the signal and estimated random noise variances, respectively, then it can be shown that $R(0)\approx \sigma_s^2 / (\sigma_s^2 + \sigma_n^2)$, so that, $1-R(0)\approx \sigma_n^2 / \overline{u'^2}$.
Since the 75\% interrogation-window overlap induces artificial correlation between neighboring vectors, only separations of three to five vector spacings were used in the fit to avoid overlap-correlated pairs. The inferred noise contribution, reported as $\sigma_n^2 / \overline{u'^2}$, was approximately 2.7\% of $\overline{u'^2}$ and 5.9\% of $\overline{v'^2}$ for the top-down measurements, and approximately 5.3\% of $\overline{u'^2}$ and 7.3\% of $\overline{w'^2}$ for the side-on dataset. 
These values indicate that the reported turbulence statistics are primarily governed by physical flow fluctuations rather than measurement noise. The larger relative noise
in the spanwise and vertical components reflects their lower fluctuation energy in the
separated region rather than poorer measurement quality, and is consistent with the
limited dynamic range of PIV when fluctuation amplitudes are small compared with the
mean particle displacement \citep{adrian2011}.


\subsection{Symmetrization \& POD\label{sec:Methods_symmetrization}}
In the following sections, the Reynolds decomposition of the velocity field is given by $\boldsymbol{u}(\boldsymbol{x},t) = \boldsymbol{U}(\boldsymbol{x})+\boldsymbol{u}'(\boldsymbol{x},t)$ where $\boldsymbol{U}(\boldsymbol{x})$ and $\boldsymbol{u}'(\boldsymbol{x},t)$ are the mean and fluctuating fields with components $U(\boldsymbol{x}),\,V(\boldsymbol{x}),\,W(\boldsymbol{x})$ and $u'(\boldsymbol{x},t),\,v'(\boldsymbol{x},t),\,w'(\boldsymbol{x},t)$, in the streamwise, spanwise and splitter-plate normal directions, $x,y,z$, respectively.

\vspace{0.5em}
\noindent
Given the existence of a symmetry plane in the mean flow (see \S\ref{sec:RD_mean}), performing POD separately on the anti-symmetric and symmetric parts of the velocity field can provide accelerated statistical convergence of the POD modes \citep{holmes2012}, while also providing clearer insight into the interactions between the anti-symmetric and symmetric motions, which are likely governed by different frequencies. Following \cite{bourgeois2013}, the symmetric/anti-symmetric split is performed on the fluctuating velocity field $(u',v')$ of the top-down $x$-$y$ plane at $z/H\,=\,0.53$, with the symmetry plane taken as $y/H=0$. Using $\boldsymbol{x}$ to denote the spatial coordinates $(x,y,z=0.53\,H)$, the velocity fluctutations can be split into the fluctuating symmetric field $\boldsymbol{u}'_{\rm{S}}(\boldsymbol{x},t)$ with components $u'_{\rm{S}},\, v'_{\rm{S}}$, and fluctuating anti-symmetric field $\boldsymbol{u}'_{\rm{A}}(\boldsymbol{x},t)$ with components $u'_{\rm{A}},\, v'_{\rm{A}}$ through the decomposition applied to the top-down plane at $z^*=0.53H$,  
\begin{equation}
    \begin{split}
        u'_{\rm{S}}(x,y,z^*,t) & = \frac{1}{2}\big( u'(x,y,z^*,t) + u'(x,-y,z^*,t) \big) \\
        u'_{\rm{A}}(x,y,z^*,t) & = \frac{1}{2}\big( u'(x,y,z^*,t) - u'(x,-y,z^*,t) \big) \\
        v'_{\rm{S}}(x,y,z^*,t) & = \frac{1}{2}\big( v'(x,y,z^*,t) - v'(x,-y,z^*,t) \big) \\
        v'_{\rm{A}}(x,y,z^*,t) & = \frac{1}{2}\big( v'(x,y,z^*,t) + v'(x,-y,z^*,t) \big).
    \end{split}
\end{equation}

\vspace{0.5em}
\noindent
Snapshot POD is then performed separately on the two fields to obtain the symmetric modes (that are symmetric in $u$ and anti-symmetric in $v$) and anti-symmetric modes (that are anti-symmetric in $u$ and symmetric in $v$). The procedure is outlined below in brief for the anti-symmetric field, but the steps are identical for the symmetric field. 

\vspace{0.5em}
\noindent
The decomposition yields a low-order reconstruction of $\boldsymbol{u}_{\rm{A}}(\boldsymbol{x},t)$ up to the leading $l$ modes:

\begin{equation}\label{eqn:pod}
    \boldsymbol{u}_{\rm{A}}(\boldsymbol{x},t) \approx \boldsymbol{U}(\boldsymbol{x}) + \sum_{i\,=\,1}^{l}\,\boldsymbol{\Phi}^{(i)}_{\rm{A}}(\boldsymbol{x})\,a^{(i)}_{\rm{A}}(t),
\end{equation}

\vspace{0.5em}
\noindent
such that $\boldsymbol{u}'_{\rm{A}}(\boldsymbol{x},t) \approx \sum_{i\,=\,1}^{l}\,\boldsymbol{\Phi}^{(i)}_{\rm{A}}(\boldsymbol{x})\,a_{\rm{A}}(t)$. The fluctuating kinetic energy of the anti-symmetric field that is captured by the first $l$ modes, $\sum_{i\,=\,1}^{l}\frac{1}{2}\overline{\big( a_{\rm{A}}^{(i)}(t) \big)^2}$, is maximized while ensuring the spatial basis functions/modes $\boldsymbol{\Phi}^{(i)}_{\rm{A}}(\boldsymbol{x})$ (with components $\Phi^{(i)}_{u,\rm{A}}(\boldsymbol{x}),\,\Phi^{(i)}_{v,\rm{A}}(\boldsymbol{x})$) are orthonormal in $\mathcal{L}^2$-Hilbert space. The time-dependent coefficients $a_{\rm{A}}^{(i)}(t)$ are orthogonal and are obtained by the orthogonal projection of $\boldsymbol{\Phi}^{(i)}_{\rm{A}}(\boldsymbol{x})$ onto $\boldsymbol{u}_{\rm{A}}'(\boldsymbol{x},t)$ with the normalization proposed by \cite{tropea}. The POD energies, $\lambda_{1,\,\rm{A}}>\lambda_{2,\,\rm{A}}>\dots,$ inform the numbering of the modes, and are eigenvalues of the two-point correlation function.

\vspace{0.5em}
\noindent
Note that the original fluctuating field can be reconstructed by adding the anti-symmetric and symmetric fields: $\boldsymbol{u}'(\boldsymbol{x},t) = \boldsymbol{u}'_{\rm{A}}(\boldsymbol{x},t) + \boldsymbol{u}'_{\rm{S}}(\boldsymbol{x},t)$. The total fluctuating kinetic energy is thus $\frac{1}{2}\sum_{i\,=\,1}^{N}\big(\lambda_{i,\,\rm{A}} + \lambda_{i,\,\rm{S}} \big)$, where $N$ is the total number of instantaneous snapshots in the dataset under consideration.

\subsection{Temporal super-resolution of low rate PIV utilizing wall pressures}

The PIV sampling rate of 15~Hz in the present study is too low to resolve the dynamics of the large-scale breathing and shedding motions (identified later in \S\ref{sec:RD_wallPressure}), limiting analysis of their temporal evolution. To overcome this limitation, a data-driven temporal super-resolution technique developed by \citet{manohar2023} is employed on the side-on datasets. The method utilizes the synchronized, time-resolved wall-pressure measurements and low-rate PIV fields, to train a long short-term memory (LSTM) neural network to predict the temporally upsampled evolution of the velocity field. Specifically, the LSTM is trained to estimate the temporal coefficients of a POD basis obtained from the undersampled PIV data, conditioned on the time history of the wall-pressure signals. The estimated coefficients are then projected onto the POD spatial modes to reconstruct high-frequency velocity fields. This approach has been shown to super-resolve 15~Hz side-on centreline PIV data of the same \textit{Gaussian Bump} flow to 2~kHz, accurately recovering the aliased large-scale breathing and shedding dynamics and the associated Reynolds-stress distributions \citep{manohar2023}.

\section{Results: Mean field \label{sec:RD_mean}}
\S\ref{sec:wallP_mean} presents the mean wall-pressure field as measured by the wall-pressure sensors. \S\ref{sec:3D_mean} highlights the 3D velocity field obtained from the measured planar PIV planes.

\subsection{Wall-pressure mean-field \label{sec:wallP_mean}}

\begin{figure}
    \centering
    \includegraphics[width=\textwidth]{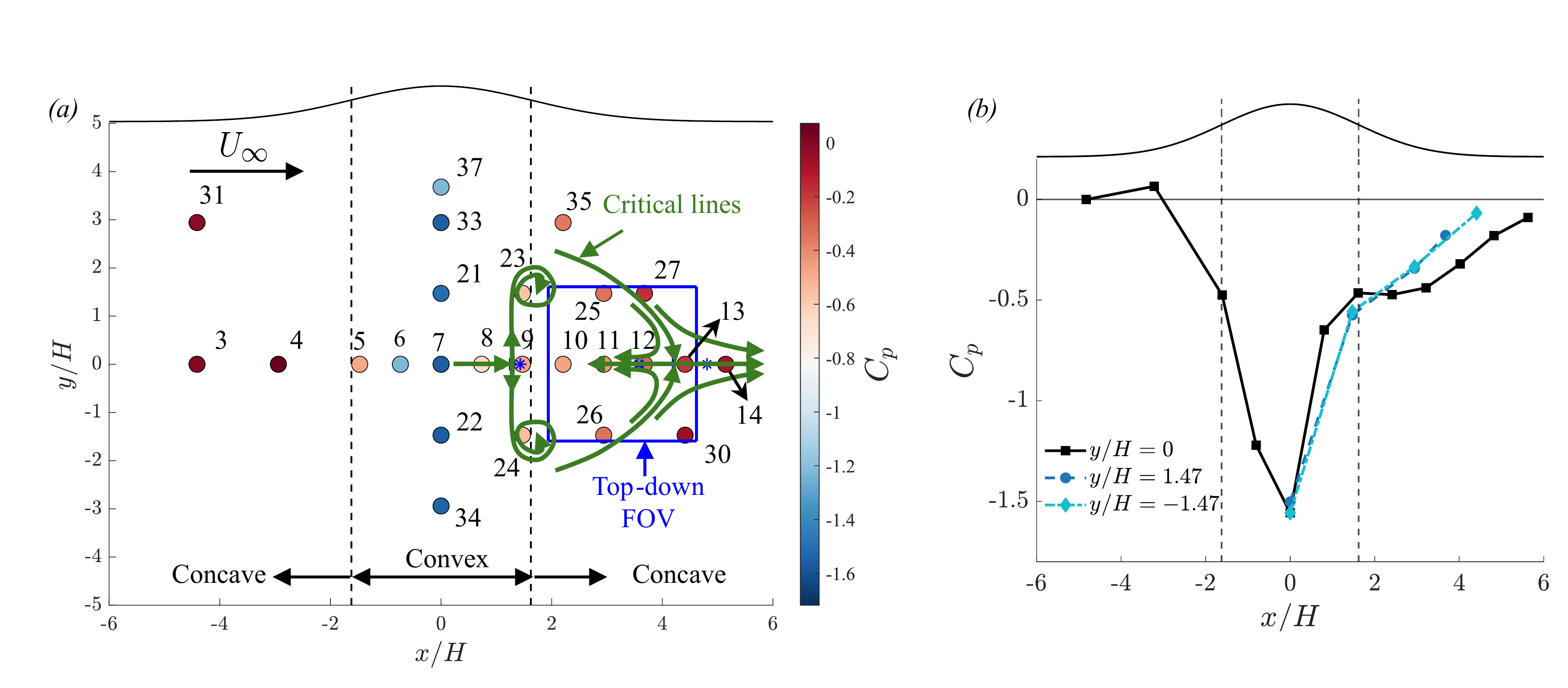}
    \caption{\small{(a) Mean $C_p$ variation over the Bump. Flow from left to right. Tap numbers labeled. Symbols along centreline $y/H=0$ represent mean (+), upstream extent (X) and downstream extent (*) of separation and downstream saddle points as determined by \citep{annamalai2022}.
    Approximate location of critical lines overlaid, based on shear-stress wall-topolopy of \cite{williams2020}.
    (b) Profiles of $C_p$ along centreline $y/H=0$ and off-centreline planes $y/H=\pm 1.47$, demonstrating mean symmetry. Lines joining dots drawn to assist visibility only. Locations of upstream and downstream surface-concavity change marked as vertical dashed lines with bump profile (top).}}
    \label{fig:Cp}
\end{figure}

The $C_p$ distribution across the 25 taps is shown in figure \ref{fig:Cp}a. The approximate location of critical points identified from flow visualization of \cite{williams2020} are overlaid in green---stable focii at $(x/H,\, y/H) \in \{(1.65,\, \approx \pm1.8)\}$ and two saddles $(x/H,\,y/H)\in\{(1.25,\,0),\,(4.2,\,0)\}$. The approximate location and range of the saddle points along the centreline, as identified by \cite{annamalai2022}, are also marked, along with the top-down PIV FOV for the present study. The $C_p$ distribution is approximately symmetric about the centreline $y/H=0$ and consistent with earlier studies \citep{robbins2021}. Profiles of the centerline $C_p$ evolution ($y/H=0$) and off-centreline planes $y/H=\pm1.47$ are highlighted in figure \ref{fig:Cp}b, showcasing this symmetry.

\vspace{0.5em}
\noindent
The interplay between surface concavity and the three-dimensionality of the downstream pressure gradient evolution is illustrated in figure \ref{fig:Cp}b. At $x/H=1.65$, the concavity of the wall changes from convex (stabilizing effect on TBL) to concave (destabilizing). In the convex region, the flow is dominated by a strong APG and separates along the spanwise front. In particular, the centreline streamwise pressure gradient ($\partial C_p/ \partial x$) rapidly relaxes until it is essentially zero for $x/H \in [1.65, \approx 2.5]$, indicating a complete loss of the APG on the centerline around the surface-concavity change. It is noteworthy that this region is dominated by the surface vortices, as illustrated by the critical lines sketched in figure \ref{fig:Cp}a.
In the concave region---for $x/H>1.65$---an APG develops again at the centreline, although not as strong as in the convex region. For $x/H > 3$, the APG appears to remain approximately constant across the span. Meanwhile, the off-centreline $\partial C_p/ \partial x$ remains positive at the location of concavity change, suggesting that the off-centreline flow 
continues to experience a streamwise APG past separation. These trends are consistent with the measurements of \citet{gray2021,gray2023} at a slightly higher Reynolds number ($\Rey_H=335\,000$). Moreover, both the present pressure measurements and prior experiments on this bump \citep{williams2020,gray2023onr} exhibit a dip in the spanwise distribution of $C_p$ at the apex ($x/H=0$) when examined along the spanwise plane $x=0$---a feature absent in 2D simulations of the same geometry. This discrepancy highlights the intrinsically 3D nature of the flow and further motivates the exploration of the unsteady motions associated with that three-dimensionality.

\subsection{3D velocity mean-field \label{sec:3D_mean}}
\begin{figure}
    \centering
    \includegraphics[width=\textwidth]{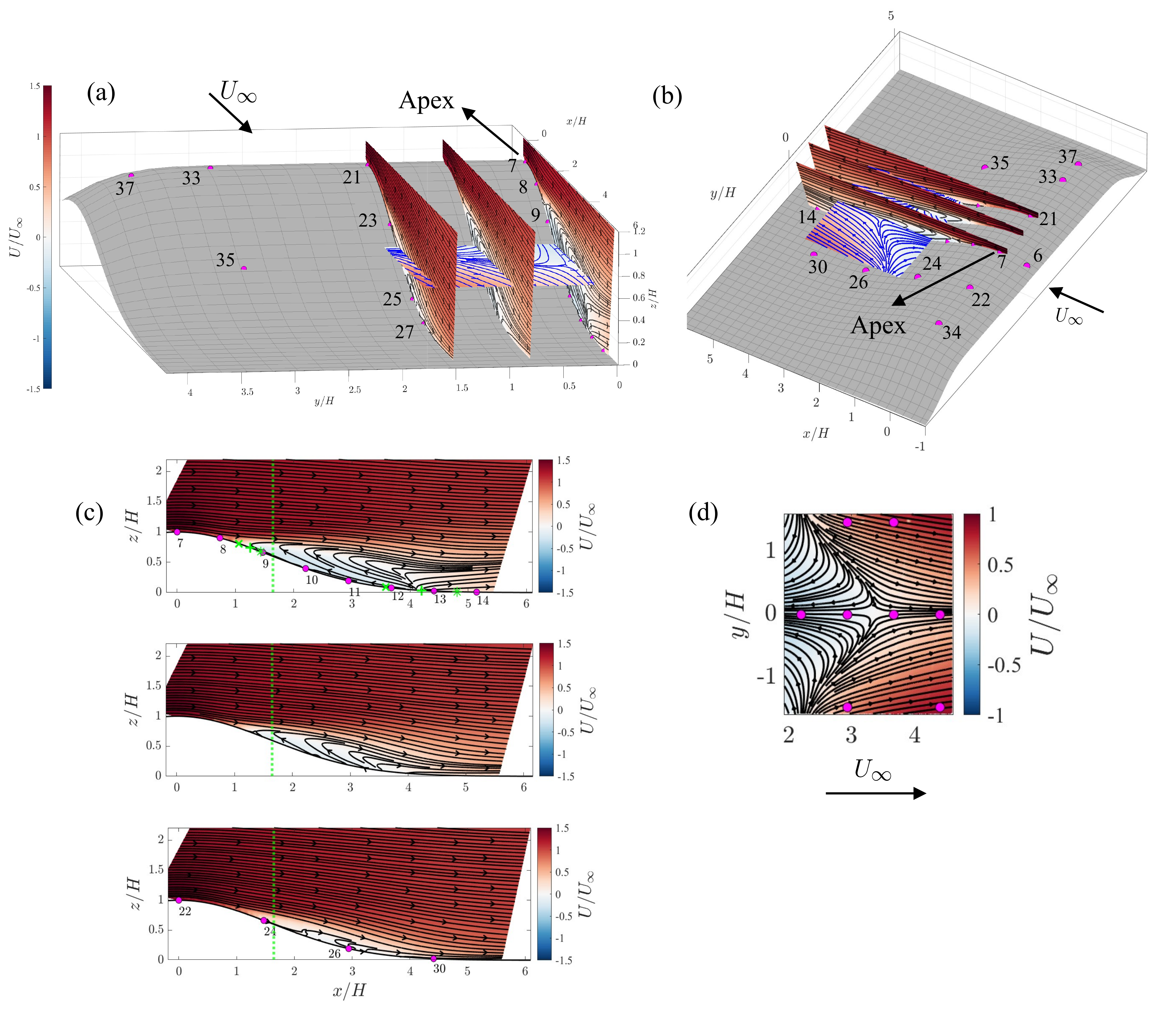}
    \caption{\small{Rendering of mean field from measurements collected at several PIV planes: (a) isometric and (b) birds-eye views. Mean fields from three overlapping FOVs along a single $x$-$z$ plane are stitched using linear interpolation to produce a planar composite mean field. Measurements are taken at (c) three separate side-on $x$-$z$ planes at $y/H\,=\,0,\,0.75,\,1.47$. Mean flow field for the (d) top-down, $x$-$y$ plane at $z/H\,=\,0.53$. Flow symmetry along centerline plane $y/H\,=\,0$. Contours represent streamwise mean velocity $U/U_{\infty}$. Magenta dots on wall represent pressure sensor locations. Select pressure taps are labeled.}}
    \label{fig:3Drendering}
\end{figure}

Aspects of the mean flow-fields measured by PIV have been extensively explored by \cite{gray2022, gray2023onr} at a slightly higher Reynolds number at \(\Rey_H=335\,000\). Here, we provide a concise summary of the present dataset at $\Rey_H = 2.26\times10^5$, highlighting its broad similarity to prior studies (e.g., \cite{gray2022, gray2023onr}) despite moderate differences in inflow conditions and Reynolds number. Representative reconstructions of the mean PIV fields relative to the bump geometry are shown in figure~\ref{fig:3Drendering}. For completeness, additional statistics---including Reynolds stresses and turbulence production---are reported in Appendix~\ref{sec:RD_flowfieldTurbulence} for comparison with existing experiments.

\vspace{0.5em}
\noindent
The flow accelerates over the apex at $(x/H,y/H)=(0,0)$, reaching a peak velocity of approximately $1.5\,U_\infty$ ($\approx 67.5$~m\,s$^{-1}$), and remains symmetric in the mean about the centreline plane ($y/H=0$).  Side-on PIV planes at $y/H = 0$, $0.75$ and $1.47$ reveal a well-defined separated region on the leeward side of the bump. Using conditional averaging based on reverse–flow fraction of the centreline side-on PIV data (see \citet{annamalai2022} for details), the upstream saddle (mean separation point) is located at $x/H=1.25$ and fluctuates over $x/H\in[1.06,\,1.43]$, while the downstream saddle lies at $x/H=4.2$ with fluctuations over $x/H\in[3.58,\,4.81]$. The corresponding mean reverse-flow length based on the Euclidean distance of mean centreline upstream and downstream saddle locations is $L_{\rm sep}\approx 0.226$~m ($=2.7H=0.21\,L$). 

\vspace{0.5em}
\noindent
These values are of the same order as \citet{gray2023onr}, who—at a higher Reynolds number ($\Rey_H=3.35\times10^5$ or $\Rey_L=4.0\times10^6$ at $U_{\infty}=69$ m/s)—report a centreline mean separation at $x=0.08L=0.95H$ and a downstream saddle\footnote{Although \citet{gray2023onr} refer to this location as a reattachment point, we label it as a downstream saddle since the mean field shows no canonical recirculation bubble consistent with true reattachment.} near $x=0.36L=4.3H$ based on mean PIV data, giving $L_{\rm sep} \approx 0.30L = 3.6H$. The slightly larger $L_{\rm sep}$ in \citet{gray2023onr} is plausibly attributable to $\Rey$ effects on shear-layer deflection---i.e., the separation zone grows with $\Rey$ \citep{williams2021}. In what follows, we adopt the present estimate of mean separation at $x = 1.25\,H \approx 0.10L$ with $L_{\rm sep}\approx2.7H=0.21\,L$ at $\Rey_H=2.26\times10^5$.

\vspace{0.5em}
\noindent
The critical-point topology observed in the side-on and top-down mean fields is consistent with the skin-friction critical points identified in figure~\ref{fig:Cp}a. In particular, the centreline side-on mean field shows that the upstream saddle coincides with the onset of mean flow separation, whereas the downstream saddle anchors a bifurcation line that extends vertically in the $x$--$z$ plane. Outboard flow descends toward the wall and converges onto this downstream saddle. Downstream of this second saddle, the flow experiences an upwash at the centreline---see the top side-on mean flow-field in figure \ref{fig:3Drendering}c. This bifurcation line terminates at a third saddle visible in the top-down field at $z/H=0.53$ (figure~\ref{fig:3Drendering}d).  
In this sense, the top-down PIV streamlines should be interpreted as planar projections of truly three-dimensional mean streamlines; care must therefore be exercised when drawing conclusions about dynamical features of the flow from projected streamline patterns alone.  
The resulting critical-line arrangement matches the well-known owl-face-of-the-first-kind topology \citep{perryHornung1984, williams2022}, consistent with oil-film interferometry measurements by \cite{gray2023onr}.

\vspace{0.5em}
\noindent
The top-down mean field highlights two dynamically distinct regions separated by a mean separatrix. Within the separated zone, the measurement plane is submerged within the reverse-flow region and mean motion is upward/outward from the plane. Outside the separatrix, the plane primarily samples fluid above the separated zone. In the mean sense, fluid is coming out of the plane within the separated zone, while fluid is going into the plane in the forward-flow regions due to the increasing wall-ward deflection of the shear layer away from the centreline.

\vspace{0.5em}
\noindent
The reverse-flow region exhibits strong three-dimensionality: its spanwise extent decreases rapidly with $|y|/H$ and the flow becomes predominantly attached for $|y|/H \gtrsim 1.5$, even though the bump shoulders lie at $|y|/H \approx 4$. This yields a separation-zone aspect ratio $B_w/L_{\rm sep} \approx 2.83$, underscoring the inadequacy of quasi-two-dimensional assumptions commonly used in simplified simulations \citep{williams2022}.

\subsubsection{Top-down flow-field turbulence characteristics}\label{sec:TD_mean_turbulence}
We examine the turbulence characteristics in the top-down plane at \(z/H=0.53\), which intersects the separated shear layer over much of its streamwise extent and provides direct access to spanwise variations within the recirculating region.

\begin{figure}
    \centering
    \includegraphics[width=\textwidth]{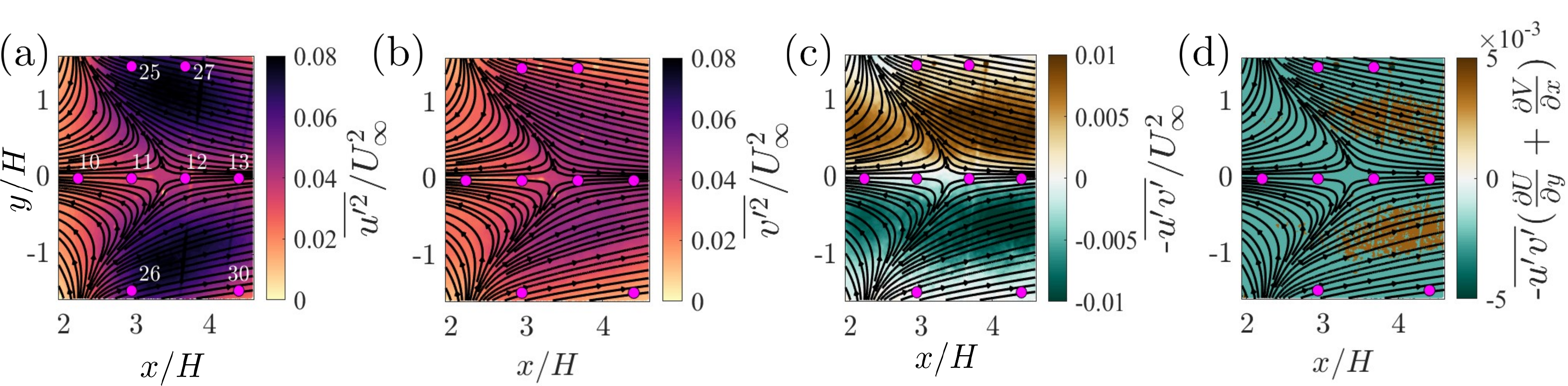}
    \caption{\small{Reynolds stresses overlaid with mean streamlines: $(a)\,\,\overline{u'^2}$ with tap numbers labelled*,\, $(b)\,\,\overline{v'^2}$,\, $(c)\,\,\overline{u'v'}$, and $(d)$ (planar) turbulence production.\\
    *Magenta dots show location of wall-pressure taps projected onto top-down plane at $z/H=0.53$.}}
    \label{fig:TD_meanStressesProd}
\end{figure}

\vspace{0.5em}
\noindent
The streamwise Reynolds stress $\overline{u'^2}$ peaks along the separated shear layer, consistent with enhanced velocity fluctuations across the high-shear interface (figure~\ref{fig:TD_meanStressesProd}a). The spanwise Reynolds stress $\overline{v'^2}$ attains its maximum magnitude near the centreline but remains approximately 40\% lower than $\overline{u'^2}$ (figure~\ref{fig:TD_meanStressesProd}b), indicating weaker spanwise fluctuations relative to the streamwise component. The Reynolds shear stress $\overline{u'v'}$ exhibits its largest magnitudes at the streamwise location where the $z/H=0.53$ plane intersects the shear layer, and the corresponding production term (figure~\ref{fig:TD_meanStressesProd}d) forms an off-centre, streamwise-elongated band aligned with regions of strong mean shear. This pattern reflects the three-dimensional deflection of the shear layer toward the wall with increasing spanwise distance from the centreline.

\vspace{0.5em}
\noindent
The regions of strongest turbulence production and Reynolds-stress anisotropy occur away from the centreline, between the pressure-tap planes, rather than along the centreline-symmetry plane itself. This spatial localization suggests that the most energetic unsteady motions are expected to originate in these off-centreline regions, a point that is borne out by the wall-pressure spectra and dominant POD modes examined in \S\ref{sec:RD_unsteady}.

Taken together, these observations establish that the present mean flow shares similar topological features reported in previous studies on slender three-dimensional bumps and hills. The separation topology, saddle locations, shear-layer deflection, and spanwise contraction of the separated region are strongly consistent across Reynolds numbers and facilities. This correspondence provides confidence that the unsteady dynamics characterized later in this paper are representative of the broader class of three-dimensional separated flows over smooth bump-type geometries.

\section{Results: Unsteady dynamics \label{sec:RD_unsteady}}
Key frequencies are identified using high-frequency wall-pressure measurements in \S\ref{sec:RD_wallPressure}, and the interaction between pairs of symmetric taps are analyzed in an effort to reveal the nature of the very low frequency motions. The associated unsteady dynamics of the spatial structures in the side-on view are discussed in \S\ref{sec:RD_Unsteadiness_xz}. The unsteady spanwise modes in the top-down view at $z/H=0.53$ are assessed in \S\ref{sec:RD_Unsteadiness_xy}.
\subsection{Separation dynamics from wall-pressure\label{sec:RD_wallPressure}}
\begin{figure}
    \centering
    \includegraphics[width=0.9\textwidth]{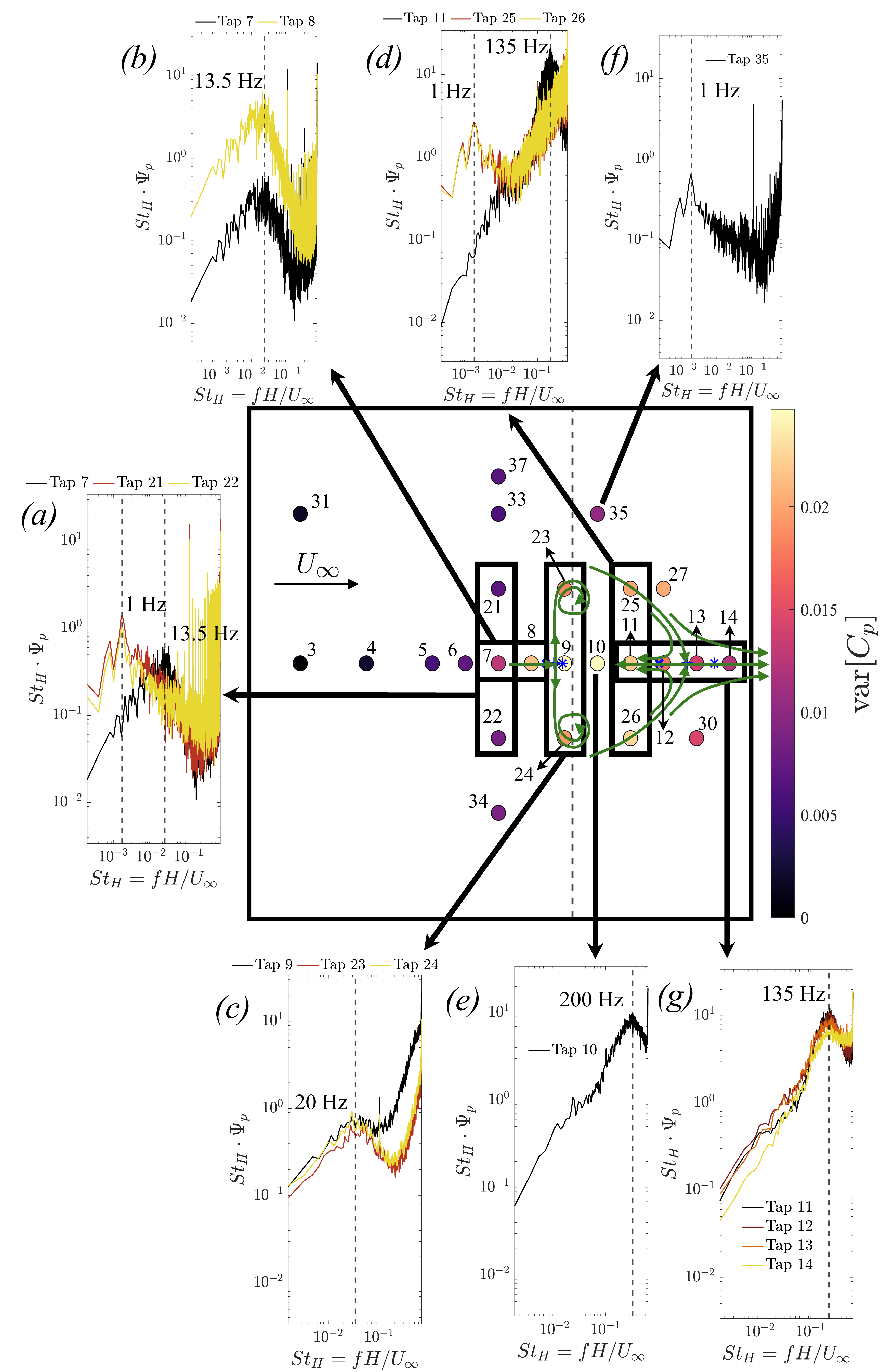}
    \caption{\small{Variance of $C_p$; refer to figure~\ref{fig:Cp}a for axes orientation. Location of downstream surface-concavity change marked as vertical dashed line. $(a-g)$ Premultiplied wall-pressure PSDFs of select (groups of) taps identified to highlight multiple-frequency phenomena within the separated zone. Centre of dominant spectral peaks marked as vertical dotted line in each inset. }}
    \label{fig:CpRMS_spectra}
\end{figure}

A spatial overview of the variance in $C_p$ as a measure of fluctuation intensity is provided in figure \ref{fig:CpRMS_spectra}, along with the pre-multiplied power spectral density (PSD) of pressure signals from select groups of taps. The fluctuation intensity of $C_p$ is low upstream of the bump, and grows considerably on the upstream slope, with the largest fluctuation intensity observed in the recirculating region---specifically near the location of surface-concavity change (see tap 10).
The distribution is close to symmetric about the centreline, just as it is in the mean. 
Wall-pressure spectra reveal several dynamically important frequencies spanning multiple orders of magnitude. In the separated region, five broad frequency bands are identified, centered at approximately 1 Hz, 13.5 Hz, 20 Hz, 135 and 200 Hz. These correspond to Strouhal numbers $St_H = fH/U_{\infty} = 1.7\times10^{-3}$, 0.023, 0.034, 0.23, and 0.34, respectively.

\vspace{0.5em}
\noindent
The 1 Hz signal is strongly expressed off-centreline (see figures \ref{fig:CpRMS_spectra}a, d, f), with the exception of locations near the surface-concavity change (taps 23/9/24).
No other very-low-frequency peaks ($f<20$ Hz) appear in these regions.
In contrast, the 13.5 Hz peak occurs only along the centreline and only upstream of mean separation and the concavity change---see figure \ref{fig:CpRMS_spectra}b. 
In particular, between taps 7 and 8 ($x/H\in[0, 0.73]$)---upstream of mean centreline separation where the APG is strongest (figure \ref{fig:Cp}b)---the spectral energy at 13.5~Hz increases by roughly an order of magnitude.

\vspace{0.5em}
\noindent
The 20 Hz peak is detected only in the tap-23/9/24 plane ($x/H=1.47$), near the surface-concavity change---figure \ref{fig:CpRMS_spectra}c. Its weaker spectral signature, coupled with the known presence of surface vortices in this region \citep{williams2020}, suggests that the 20~Hz is likely associated with their motion. The higher-frequency content at centreline tap 9 compared to off-centreline taps 23/24 may indicate the onset of small-scale shear-layer structures.

\vspace{0.5em}
\noindent
The highest frequency spectral peaks (135 Hz, 200 Hz) occur downstream of separation along the centerline, where vortex shedding is expected to occur---figures \ref{fig:CpRMS_spectra}(e, g). The 200 Hz peak only occurs immediately after separation (tap 10), while the 135 Hz peak emerges thereafter and persists  across taps 11-14.
This progression suggests a redistribution of spectral energy from higher to mid-frequencies along the separated shear layer. As noted by \cite{manohar2023}, the decrease from 200 Hz to 135 Hz may reflect vortex amalgamation during downstream advection, a phenomenon often observed in separated flows \citep{cherry1984}. The gradual weakening of the 135 Hz peak farther downstream may indicate that vortices are moving away from the wall as they are advected downstream. This is supported by the centreline upwash observed in the side-on mean PIV fields (figure \ref{fig:3Drendering}c, top).

\begin{figure}
    \centering
    \includegraphics[width=\textwidth]{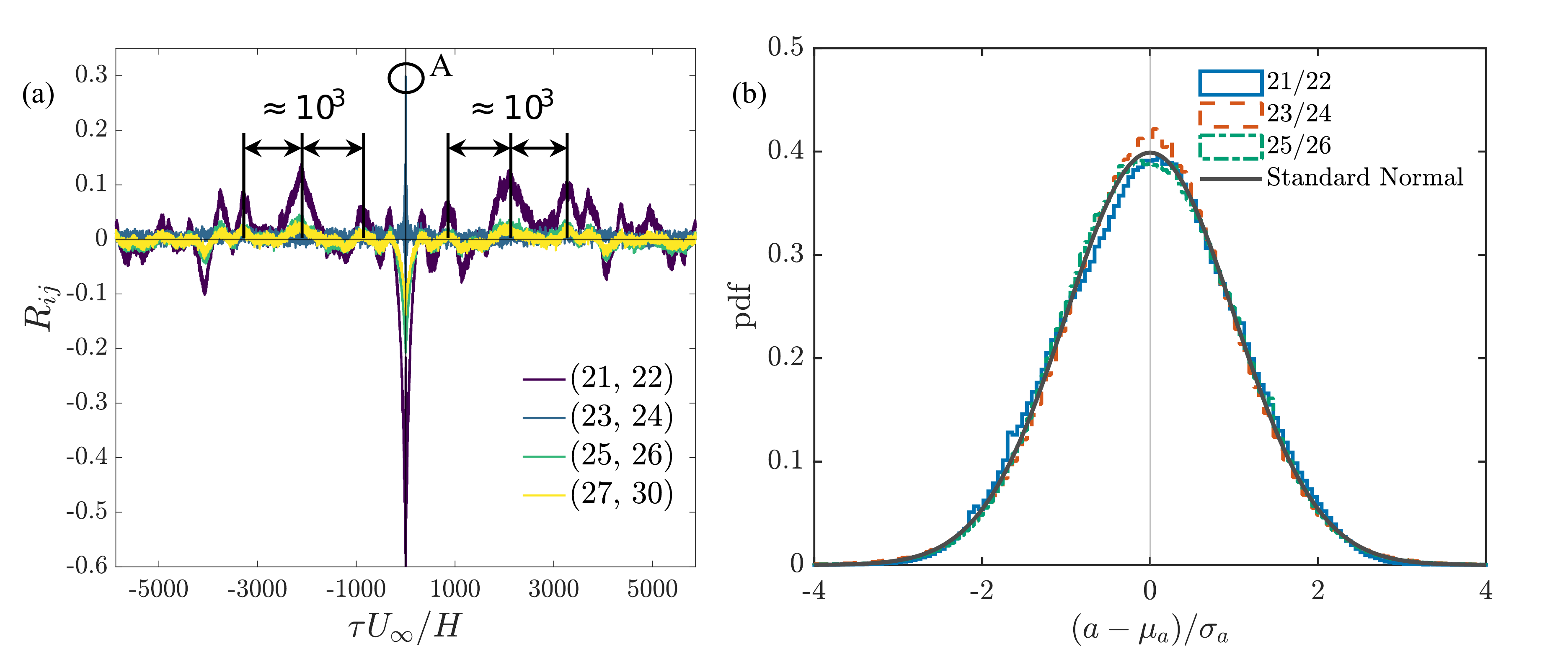}
    \caption{\small{(a) Cross-correlation of fluctuating surface-pressure signals across centreline symmetry plane $y/H=0$. Point A marks the maximum (positive) correlation between Taps 23/24. Pressure signals between taps are correlated in the order presented in parenthesis shown in the legend. Signals low-pass filtered with 400 Hz cut-off, and resampled at 800 Hz. (b) Normalized PDFs of anti-symmetric field across symmetric tap pairs, compared with the standard normal distribution. }}
    \label{fig:xcorr}
\end{figure}

\vspace{0.5em}
\noindent
Cross-correlations between wall-pressure signals measured at various taps provide insight into the nature of the coherent motions across the bump. The cross-correlation between discrete fluctuating pressure signals pertaining to two taps, $p'_i(t)$ and $p'_j(t)$, is defined as,
\[R_{ij}(\tau)= \overline{p'_i(t)\,p'_j(t + \tau)}\,\,/\,\,\sqrt{\overline{p'^2_i(t)} \cdot \overline{p'^2_j(t)}},\]
as shown in figure \ref{fig:xcorr} for pressure tap pairs either side of the centerline.

\vspace{0.5em}
\noindent
Interestingly, taps~23 and~24 are the only pair exhibiting a positive correlation at zero time lag (marked `A'), and they lie closest to the estimated locations of the surface–vortex cores.  
Their proximity to these core regions supports the earlier inference that the $\sim 20$~Hz signature arises from the dynamics of the surface vortices, because this frequency band is expressed most strongly at these taps. Meanwhile, the positive cross-correlation across the centreline indicates that the associated pressure motions are predominantly in-phase on either side of the bump, consistent with the symmetric configuration of the surface vortices as illustrated in figure \ref{fig:Cp}a.

\vspace{0.5em}
\noindent
Negative correlations at zero time-lag are observed across the centreline along the other streamwise locations---taps 21/22, 25/26, and 27/30---where the 1 Hz motion ($St_H \approx 10^{-3}$) is observed. Interestingly, these taps exhibit a periodicity in the cross-correlation with longer periods spanning $10^{3}$ $H/U_{\infty}$ convective time units. The periodicity is most pronounced between taps 21/22 (off-centerline bump apex). These features suggest that the lowest-frequency motion at 1 Hz is a side-to-side oscillation of the separation region that is anti-symmetric relative to the centreline.

\vspace{0.5em}
\noindent
The nature of this oscillation remains to be determined---i.e., whether it is meandering/harmonic or more bi-modal, like the wakes of the Ahmed body  \citep{grandemange2013}, 3D axisymmetric hill  \citep{byun2006,garcia2009} and BeVERLI hill \citep{gargiulo2021,duetsch2022,macgregor2023}. 
To assess potential bi-modality, the anti-symmetric component of the wall-pressure field is taken across the centreline symmetry plane: $a(t)=\frac{1}{2}(p_2(t) - p_1(t))$, where $p_1$ and $p_2$ are pressure signals corresponding to sensors pairs on opposite sides of the centreline. It is computed for the tap pairs 21/22, 23/24 and 26/27, and the z-score normalized probability density functions (PDFs) are plotted in figure \ref{fig:xcorr}b. The PDFs are found to be approximately normally distributed across the entire separated wake. Therefore, unlike the bi-stable wakes, the VLF spanwise motion ($St_H = 10^{-3}$) appears to be a continuous oscillation, or a meandering about a mean state\footnote{It is also noted that the $r$-parameter proposed by \cite{plumejeau2019}---$ r(t) = \frac{p_2(t) - p_1(t)}{p_2(t) + p_1(t)}$---was not used for the bi-modality analysis as taps 25/26 contain several events where $p_2\approx -p_1$, causing $r$ to become unbounded.}.

\subsection{Characteristics of the breathing and vortex shedding modes observed from side-on flowfields}\label{sec:RD_Unsteadiness_xz}

\begin{figure}
    \centering
    \includegraphics[width=\textwidth]{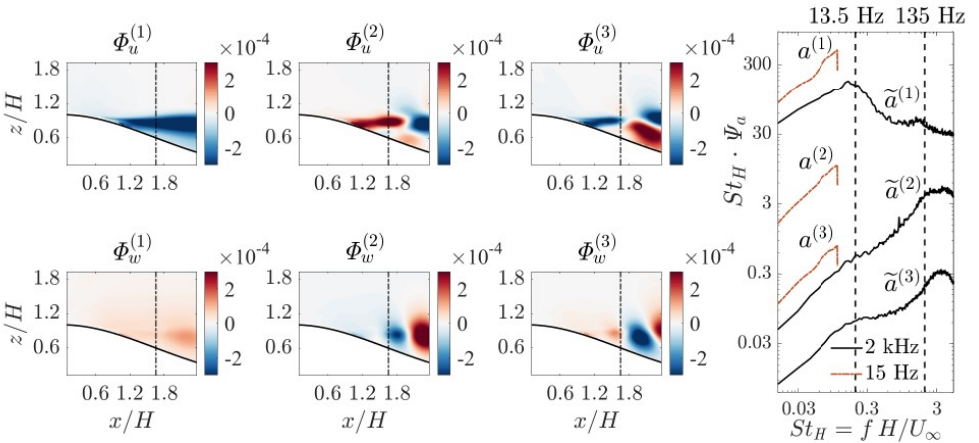}
    \caption{\small{[Adapted from \citet{manohar2023}]. Columns 1-3 (from left to right) correspond to the first three spatial POD modes with $u$-(top) and $w$-components (bottom). Column 4 shows pre-multiplied PSDs of corresponding original $a^{(1-3)}$ (15 Hz) and super-resolved $\widetilde{a}^{(1-3)}$ (2 kHz) coefficients estimated from unseen pressure inputs spanning [the test dataset]. The PSDs of each mode are offset by $10^{-1}$ for clarity. Modes 1-3 represent 33.8\%, 7.5\% and 5.2\% of the planar turbulent kinetic energy, respectively. }}
    \label{fig:EXIF_modes}
\end{figure}

It is well-known that the dynamics of the shear layer at high Reynolds number are dominated by advecting large-scale vortical structures emerging from Kelvin-Helmoltz instabilities at separation \citep{simpson1989,cherry1984}. To explore the physical sources of the 13.5 and 135-200 Hz frequencies, figure \ref{fig:EXIF_modes} surveys the frequencies of the dominant PIV modes in side-on PIV fields at the bump centerline. The method of \cite{manohar2023} has been used to upsample the low-rate PIV fields using unsteady wall pressure signals such that otherwise-aliased frequencies can be investigated. The first three spatial POD modes $\Phi^{(1-3)}$ with components $\it{\Phi^{(1-3)}_u},\,\it{\Phi^{(1-3)}_w}$ are shown, along with their super-resolved temporal coefficients, $\widetilde{a}^{(1-3)}$. Following \cite{manohar2023}, the dynamics of the first POD mode are dominated by a 13.5 Hz spectral signature that was observed strictly upstream of mean centreline separation in the wall-pressure signals. 

\begin{figure}
    \centering
    \includegraphics[width=\textwidth]{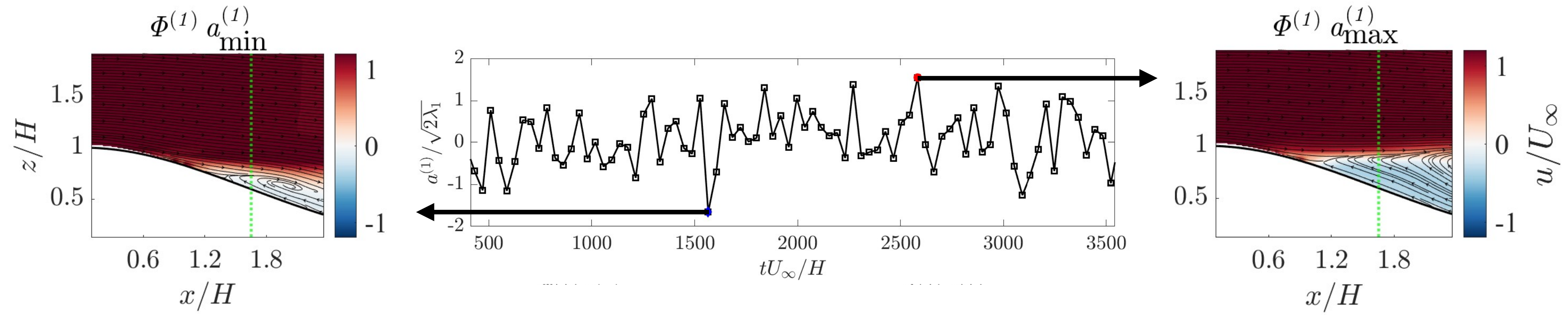}
    \caption{\small{Time-series of breathing mode coefficient $a^{(1)}$ with labelled local minimum and maximum, with corresponding mode-1 reconstructions. Dots connected by lines to assist visibility only.}}
    \label{fig:minmax_a1}
\end{figure}

\vspace{0.5em}
\noindent
Single-mode reconstructions of the super-resolved mode-1 overlaid with the mean field (figure \ref{fig:minmax_a1}) reveal its association with the motion of the separation point. This motion modulates the instantaneous extent of the reverse-flow region, resulting in a vertical motion of the shear layer, or `breathing' motion of the separation zone.
The local minimum $a^{(1)}_{\text{min}}<0$ appears to correspond to a flow state with a smaller reversed-flow region, while the local maximum $a^{(1)}_{\text{max}}>0$ corresponds to a state with a larger separation zone. The separation point appears to move correspondingly (see Supplementary file 2 of \cite{manohar2023}).  The intermediate states corresponding to $a^{(1)}_{\text{min}}<a^{(1)}<a^{(1)}_{\text{max}}$ describe a range of flow states associated with the breathing mode. The above results are consistent with the breathing behavior of the pressure-induced TSB flows studied by \cite{taifour2016, wu2020, fang2024}.

\vspace{0.5em}
\noindent
The spatial signatures of modes~2 and~3, along with the super-resolved spectra of their temporal coefficients, show that these modes represent the formation, amalgamation, and downstream advection of large-scale coherent structures tied to vortex-shedding dynamics in the $135$–$200$~Hz band. We interpret these structures as originating from nascent Kelvin–Helmholtz instabilities along the separated shear layer—whose smallest emerging vortices populate the upper part of this band ($\sim\!200$~Hz)—followed by nonlinear vortex amalgamation and advection that redistribute the energy toward the observed $\sim135$~Hz peak (figure~\ref{fig:EXIF_modes}).

\subsection{Characteristics of spanwise motions}\label{sec:RD_Unsteadiness_xy}
The top-down flow field in the separated region at $z/H = 0.53$ is analyzed by splitting the POD basis into antisymmetric (A) and symmetric (S) families with respect to the centreline, allowing the dynamics of spanwise structures in the separated region to be explored. 
Coefficients of each of these families are written $a^{(i)}_{\rm{A}}$ and $a^{(k)}_{\rm{S}}$, where $i,k$ are mode numbers. Unless otherwise stated, all POD coefficients are presented in their normalized form, $a/\sqrt{2\,\lambda}$, with $\lambda$ denoting the corresponding modal energy.

\begin{figure}
    \centering
    \includegraphics[width=\textwidth]{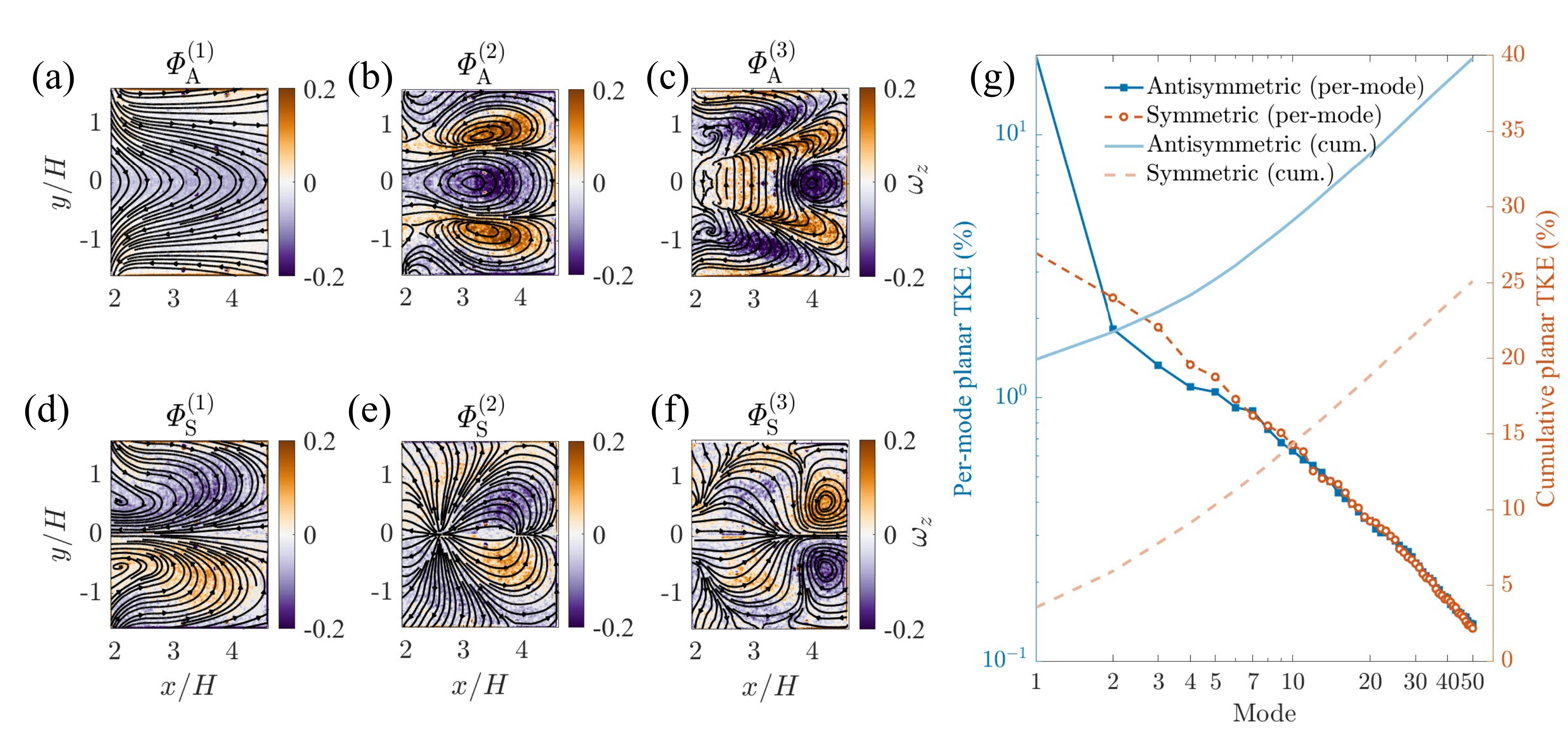}
    \caption{\small{POD mode shapes and energies. (a-c) First three antisymmetric modes $\boldsymbol{\Phi}_{\rm{A}}^{(1-3)}$ and, (d-f) first three symmetric modes $\boldsymbol{\Phi}_{\rm{S}}^{(1-3)}$. Contours represent vertical vorticity component $\omega_z$; streamlines drawn from corresponding modal velocity field. (g) Modal energy distribution - per-mode planar TKE (left, log scale) and cumulative fraction (right, linear) - for antisymmetric (solid) and symmetric (dashed) families.}}
    \label{fig:SA_modes_modalE}
\end{figure}

\vspace{0.5em}
\noindent
The first three antisymmetric and symmetric spatial modes are shown in figure \ref{fig:SA_modes_modalE}. The most energetic mode is the leading antisymmetric mode, \(a_{\text{A}}^{(1)}\), contributing to around 20\% of the total planar fluctuating kinetic energy (see figure \ref{fig:SA_modes_modalE}g).  In contrast, the leading symmetric mode, \(a_{\text{S}}^{(1)}\), contains 3\% of the planar fluctuating energy. The higher-order antisymmetric modes contain less energy than \(a_{\text{S}}^{(1-6)}\), and are thus disregarded in the following analysis.

\begin{figure}
    \centering
    \includegraphics[width=0.7\textwidth]{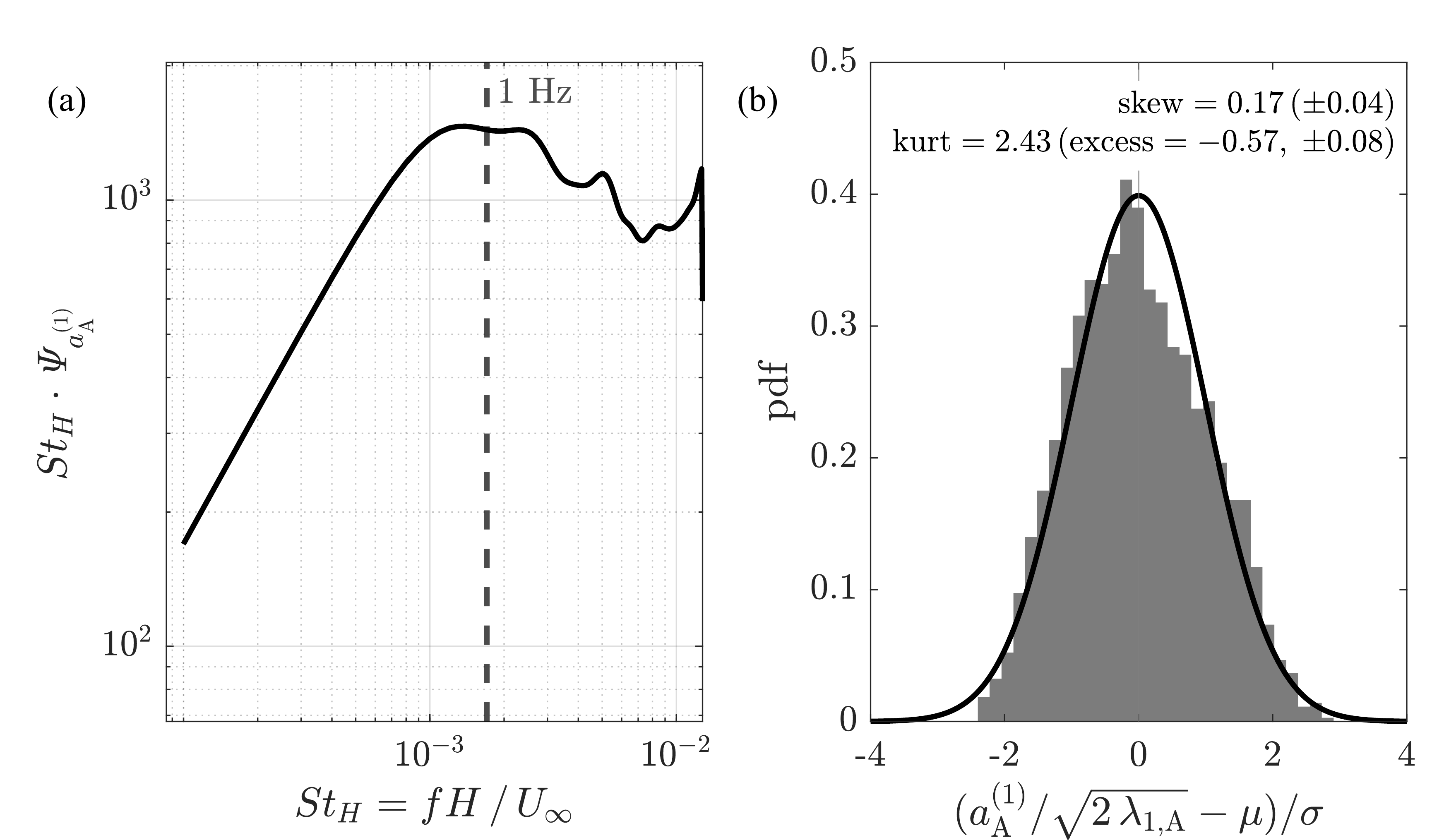}
    \caption{\small{(a) Premultiplied spectrum of the antisymmetric mode-1 coefficient; vertical dashed line marks the 1 Hz peak. (b) Probability density of the standardized coefficient with a standard normal reference (solid line); annotations report skewness $0.17 \pm 0.04$ and kurtosis 2.43 (excess $-0.57 \pm 0.08$).}}
    \label{fig:a1_psd_pdf}
\end{figure}

\vspace{0.5em}
\noindent
The dynamics of \(a_{\text{A}}^{(1)}\) are summarized in figure \ref{fig:a1_psd_pdf}. The pre-multiplied PSD of $a^{(1)}_{\rm{A}}$ shows a broadband hump centred at 1 Hz, or $St_H=fH/U_{\infty}\approx10^{-3}$, consistent with the off-centreline wall-pressure spectra reported in \S\ref{sec:RD_wallPressure}. Note that the 1 Hz-centered spectral hump was absent from the coefficients of all modes (antisymmetric and symmetric) other than \(a_{\text{A}}^{(1)}\).
The standardized PDF of $a^{(1)}_{\rm{A}}$ is near-Gaussian (skew=0.17), with significantly lighter-than-normal tails (excess kurtosis = -0.57), indicating weak asymmetry and low intermittency---consistent with a low-frequency meandering about the mean, i.e., a continuous lateral wake oscillation without long residence in a preferred state. This matches the conclusion from the bi-modality analysis conducted on the wall-pressure signals in \S\ref{sec:RD_wallPressure}, that the $St_H\approx10^{-3}$ spanwise motion is a meandering mode, which is in stark contrast to the bi-stable wakes observed in other geometries, where the wake prefers more extreme conditions away from the mean, with reasonably long residence times in theses states. 

\begin{figure}
    \centering
    \includegraphics[width=\textwidth]{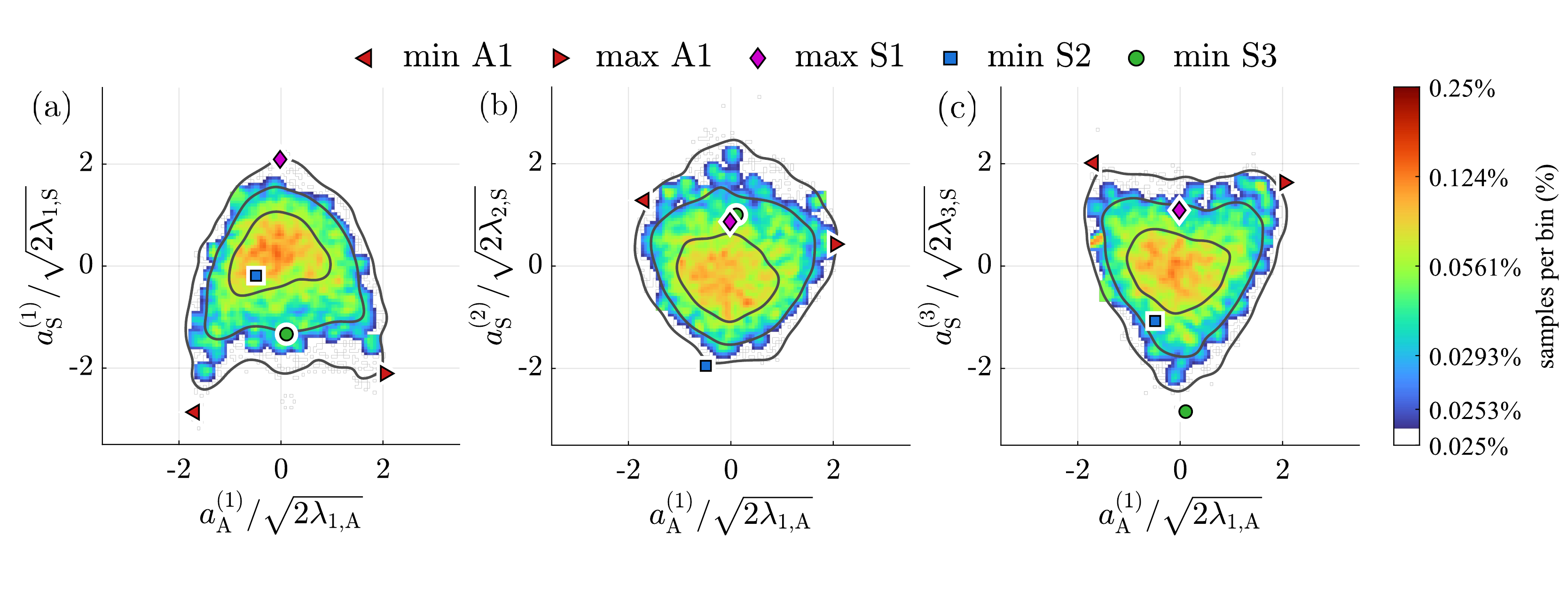}
    \caption{\small{Joint distributions of POD coefficients comparing \(a_{\text{A}}^{(1)}\) with \(a_{\text{S}}^{(1-3)}\). Each panel shows the 2D histogram (120 x 120 bins) of $(a_{\text{A}}^{(1)},\,a_{\text{S}}^{(i)})$ after normalization. Color indicates fraction of samples per bin. Black contours denote highest-density regions enclosing 50\%, 90\% and 99\% of the probability. Select extreme (minimum/maximum) points marked as colored symbols; the position of a given extreme point in any of the three distributions corresponds to the same time instant. Axes are equal with common limits. 
    }}
    \label{fig:keyPoints_jointPDF}
\end{figure}

\vspace{0.5em}
\noindent
To probe the coupling of dominant spanwise antisymmetric and symmetric motions, the joint distributions of normalized coefficients $(a^{(1)}_\text{A}, a^{(i)}_\text{S}),\,i=1:3$ are examined in figure \ref{fig:keyPoints_jointPDF}.
It is observed that probability distributions are centered  near $a_{\rm{A}}^{(1)} = 0$, showing that the flow spends most of the time close to the symmetric mean position. Interestingly, the structure of the distributions suggests an underlying coherence in the inter-modal dynamics. Two trends emerge:
\begin{itemize}
    \item In figure \ref{fig:keyPoints_jointPDF}a, \(\{a_{\text{A}}^{(1)},\,a_{\text{S}}^{(1)}\}\) displays a parabolic structure in the low-probability rims. The distribution is nearly symmetric about $a^{(1)}_{\rm{A}}=0$: \(a_{\text{S}}^{(1)}\) is largest when \(a_{\text{A}}^{(1)}\approx0\) and becomes negative for large \(|a_{\text{A}}^{(1)}|\). 
    \item In figure \ref{fig:keyPoints_jointPDF}c,  \(\{a_{\text{A}}^{(1)},\,a_{\text{S}}^{(3)}\}\) also displays a parabolic structure in the low-probability rims, where \(a_{\text{S}}^{(3)}\) increases with \(|a_{\text{A}}^{(1)}|\), with the largest \(a_{\text{S}}^{(3)}\) events occurring at the farthest off-centreline positions.
\end{itemize}

\vspace{0.5em}
\noindent
Practically, this suggests that large \(|a_{\text{A}}^{(1)}|\) tends to co-occur with negative \(a_{\text{S}}^{(1)}\) and positive \(a_{\text{S}}^{(3)}\), noting that the signs of the POD modes are arbitrary and may flip depending on the specific realization of the decomposition. Moreover, the extreme (min/max) events lie on low-probability rims of the distributions, confirming that large spanwise excursions are rare compared with the near-symmetric state, thereby indicating the stability of the mean field.

\vspace{0.5em}
\noindent
The distribution involving $a_{\text{S}}^{(2)}$ (figure~\ref{fig:keyPoints_jointPDF}b) does not exhibit a clear geometric trend in its low-probability structure. Unlike those involving $a_{\text{S}}^{(1)}$ and $a_{\text{S}}^{(3)}$, which show a relatively well-defined parabolic structure, the $\{a_{\text{A}}^{(1)}, a_{\text{S}}^{(2)}\}$ distribution remains comparatively isotropic and lacks an unambiguous modal relation. We therefore do not extract further dynamical interpretation from this pairing.

\vspace{0.5em}
\noindent
Parabolic low-probability loci in the joint PDFs have also been reported for axisymmetric wakes \citep[e.g., cylinder with an elliptic nose cone; figure~18\textit{e} of][]{pavia2019}, where they become especially pronounced when the long-time dynamics are isolated. In that case, the parabolic structure arises from a non-linear coupling between the antisymmetric modes (azimuthally asymmetric, \(m=\pm1\)) and the symmetric deformation mode (axisymmetric bubble-pumping, \(m=0\)) during azimuthal meandering. This behavior is directly analogous to the meandering dynamics observed here in a reflectionally symmetric bump.

\begin{figure}
    \centering
    \includegraphics[width=\textwidth]{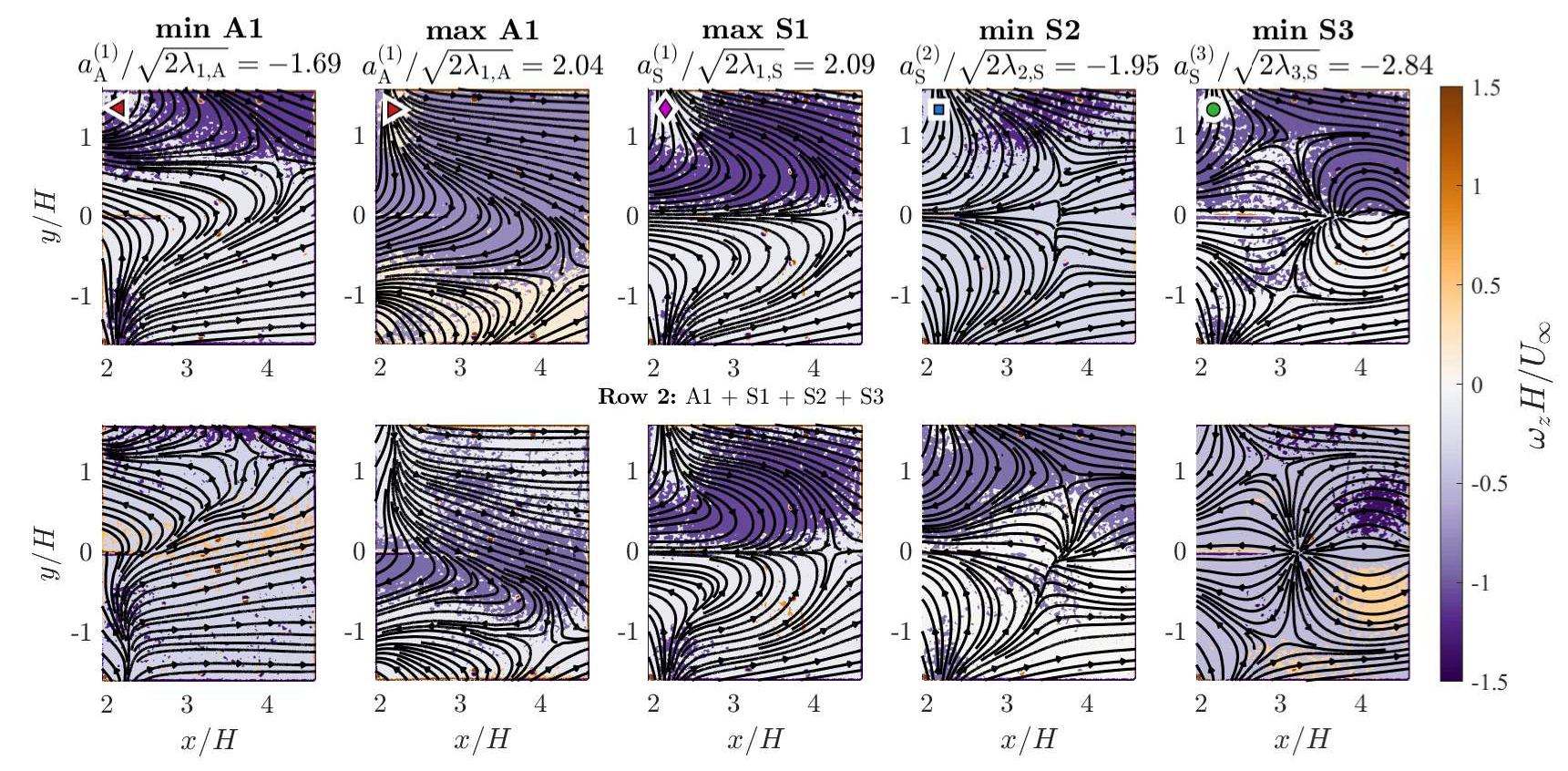}
    \caption{\small{Instantaneous flow-field reconstructions at key modal events identified in figure \ref{fig:keyPoints_jointPDF}. Columns show (from left): \textbf{min A1, max A1, max S1, min S2, min S3} added to the mean field. \textbf{Top row:} single-mode-$k$ snapshots $\boldsymbol{U}+\boldsymbol{\Phi}_{\rm{A/S}}^{(k)}\,a_{\rm{A/S}}^{(k)}$ at the indicated time; titles report the normalized coefficient. \textbf{Bottom row:} multi-mode reconstructions at the same time steps, $\boldsymbol{U}+\boldsymbol{\Phi}_{\rm{A}}^{(1)}a_{\rm{A}}^{(1)}+\sum_{i=1}^3\boldsymbol{\Phi}_{\rm{S}}^{(i)}a_{\rm{S}}^{(i)}$. Color maps vertical vorticity component $\omega_z$; black curves are instantaneous streamlines. Markers in upper-left of top panels identify event type and use the same symbols as figure \ref{fig:keyPoints_jointPDF}.}}
    \label{fig:keyPointsFields}
\end{figure}

\vspace{0.5em}
\noindent
Instantaneous flow-fields at five key modal events are selected from the joint distributions---\textit{`min A1'}, \textit{`max A1'}, \textit{`max S1'}, \textit{`min S2'}, and \textit{`min S3'}. These events correspond to time instants where \(a_{\text{A}}^{(1)}\) is a global minimum/maximum within the time-series of the dataset (\textit{`min/max A1'}, respectively), and similarly \textit{`min S1'} is the time instant corresponding to when \(a_{\text{S}}^{(1)}\) is a global minimum, and so on. Figure \ref{fig:keyPointsFields} shows flow-field reconstructions at these modal events.
The single-mode panels (top row) illustrate the effect of each mode added to the mean, whereas the second row displays four-mode reconstructions based on modes A1 and S1-S3.

\vspace{0.5em}
\noindent
Before examining the individual S1–S3 events in detail, we note that the single–mode reconstructions in figure~\ref{fig:keyPointsFields} (top row) capture the essential spatial features observed in the four–mode reconstructions of figure~\ref{fig:keyPointsFields} (bottom row). At the instants corresponding to the extreme events, the dominant symmetric or antisymmetric mode carries most of the energy, such that adding the remaining modes modifies the field only incrementally. This indicates that the extreme events are largely governed by the leading mode in each symmetry class, and therefore the following discussion focuses on the physical interpretation of these dominant deformations rather than on differences between the single–mode and multi–mode reconstructions.

\vspace{0.5em}
\noindent
Instants of \textit{`min/max A1'} appear to correspond to asymmetric wake states with the downstream saddle close to one bump height, $H$, from the centerline (figure \ref{fig:keyPointsFields}; \textit{cols.} min/max A1). Thus, the effect of adding the spanwise-meandering mode \(a_{\text{A}}^{(1)}\) to the mean field is to shift the separated region off-centreline, creating a side-to-side spanwise meandering-type oscillation at $St_H\approx10^{-3}$ (see \textbf{Supplementary Movie 1: A1}). 

\vspace{0.5em}
\noindent
The \textit{`max S1'} event coincides with an $a^{(1)}_{\rm{A}}=0$ event (see figure \ref{fig:keyPoints_jointPDF}a). In particular, \textit{`max S1'} corresponds to the event where the saddle point is positioned farthest downstream within the FOV (roughly one $H$ downstream from the mean saddle location), consistent with a symmetric streamwise-elongation of the wake (figure \ref{fig:keyPointsFields}; \textit{col.} max S1). 
Moreover, the \textit{`min/max A1'} events align with the minima of $a_{\rm S}^{(1)}$, indicating that asymmetric wake states preferentially coincide with upstream-shifted configurations of the separation region (\textbf{Supplementary Movie~2: S1}). Thus, $a_{\rm S}^{(1)}$ may be interpreted as a proxy for the streamwise extent of the separation zone. In particular, for the current decomposition, $a_{\rm S}^{(1)}>0$ corresponds to a downstream-shifted wake (relative to the mean), whereas $a_{\rm S}^{(1)}<0$ indicates an upstream contraction.

\vspace{0.5em}
\noindent
The \textit{`min S2'} and \textit{`min S3'} events also both occur when the wake is close to the symmetric position ($a^{(1)}_{\rm{A}}\approx0$)---see figure~\ref{fig:keyPoints_jointPDF}.
At large spanwise excursions ($|a^{(1)}_{\rm{A}}|$ large), the symmetric response is not limited to \(a_{\text{S}}^{(1)}\): while \(a_{\text{S}}^{(1)}\) carries the largest energy among symmetric modes, the joint distributions show that \(|a_{\text{S}}^{(2)}|\) and \(|a_{\text{S}}^{(3)}|\) likewise become appreciable. In particular, for this decomposition, $a_{\rm S}^{(1)}$ follows becomes increasingly negative with 
$|a_{\rm A}^{(1)}|$ implying a contracted wake, $a_{\rm S}^{(3)}$ increases with $|a_{\rm A}^{(1)}|$, and $|a_{\rm S}^{(2)}|$ departs from zero as the antisymmetric deformation grows. Thus, asymmetric states are generally accompanied by a contracted wake (\(a_{\text{S}}^{(1)}<0\))---at most roughly $0.5\,H$ upstream from the mean saddle location---as well as appreciable higher-order symmetric content.

\vspace{0.5em}
\noindent
The \textit{`min S2'} and \textit{`min S3'} events also produce qualitatively similar symmetric deformations in the reconstructed fields (figure~\ref{fig:keyPointsFields}, bottom row). In particular, within the top-down FOV, \textit{`min S3'} corresponds to the most upstream configuration of the \textit{symmetric} wake (green circle in figure~\ref{fig:keyPoints_jointPDF}a), whereas the largest \(a_{\text{S}}^{(3)}\) amplitudes co-occur with the extreme \(|a_{\text{A}}^{(1)}|\) states (red triangles in figure~\ref{fig:keyPoints_jointPDF}c). When modes A1+S1+S2+S3 are superposed (i.e., $\boldsymbol{U}+\boldsymbol{\Phi}_{\rm{A}}^{(1)}a_{\rm{A}}^{(1)}+\sum_{i=1}^3\boldsymbol{\Phi}_{\rm{S}}^{(i)}a_{\rm{S}}^{(i)}$) at the \textit{`min S2'} or \textit{`min S3'} instants, the reconstructed flow exhibits a split of the central saddle point (figure \ref{fig:keyPointsFields}, bottom row, \textit{cols.} min S2/S3). A similar splitting tendency is already visible in the S3 mode alone (figure \ref{fig:SA_modes_modalE} and \textbf{Supplementary Movie 3: S3}). These observations suggest that the most upstream \textit{symmetric} state is associated with a saddle-split topology---potentially an unstable intermediate state preceding spanwise wake shifting. Time-resolved measurements would be required to verify this by resolving trajectories in $(a^{(1)}_{\rm{A}}, a^{(1)}_{\rm{S}}, a^{(3)}_{\rm{S}})$ space.

\vspace{0.5em}
\noindent
A closer inspection of the multi–mode reconstructions at the \textit{`min S2'} and \textit{`min S3'} instants (bottom row of columns 4–5 in figure~\ref{fig:keyPointsFields}) reveals a topology that is distinct from the owl–face-of-the-first-kind found in the mean skin-friction field (recall figure \ref{fig:Cp}a). In both cases, the instantaneous streamline pattern in the $z/H=0.53$ plane is characterized by a centreline focal point flanked by two off–centreline saddles. This topology is locally similar to the downstream portion of the surface-shear-stress pattern of an owl–face-of-the-second-kind in the sense of \citet{perryHornung1984}, which is associated with the presence of loop vortices emanating from the wall. This instantaneously observed saddle-split topology may therefore be associated with the shedding of large loop vortices.
\section{Discussion \label{sec:discussion}}
The results above point to distinct sources for the four energetic frequency bands observed in this flow. Here we compare the measured Strouhal numbers with values reported for other separated flows (narrower hills, diffusers and axisymmetric wakes), and use these comparisons—together with the present POD, wall-pressure and modal reconstruction evidence—to propose a coherent picture of the three-dimensional dynamics that underlie these frequencies.

\vspace{0.5em}
\noindent
Table~\ref{tab:keyFrequencies} summarizes the four frequencies using several length-scale normalizations. Prior studies on square-back and axisymmetric bodies often use an axis-agnostic scale (e.g.\ base height $H$ or diameter $D$), which is reasonable for unit aspect-ratio geometries ($B_w/H=1$). The \textit{Gaussian Bump}, however, is wide ($B_w/H\approx 8.3$), the separated region is elongated in the streamwise direction ($L_{\rm sep}\approx 3H$ at $y/H=0$), and the underlying mechanisms that result in each of the identified frequencies may most naturally scale with different lengthscales.  
We therefore report $St_H=fH/U_\infty$, $St_{L_{\rm sep}}=fL_{\rm sep}/U_\infty$, and $St_{B_w}=fB_w/U_\infty$ so that comparisons can be made with a range of results from the literature.

\begin{table}
  \centering
  \begin{tabular}{ccccc}
    $f$ (Hz) & $St_{H}$ & $St_{L_{\rm sep}}$ & $St_{B_w}$ & Dominant location / taps\\
    \hline
     1        & \textbf{0.0017} & 0.0050 & 0.014  & off-centreline taps 21–27, 30, 33–37 \\
     13.5     & 0.023           & \textbf{0.068} & 0.19   & centreline upstream of separation (taps 7–8) \\
     20       & 0.034           & 0.10   & {0.28} & near surface vortex cores (taps 7/23/24) \\
     135–200  & 0.23-0.34   & \textbf{0.68-1.01}   & 1.9-2.81    & centreline shear layer (taps 11–14) \\
  \end{tabular}
  \caption{\small Key frequencies and Strouhal numbers. The most physically relevant normalization is bolded for each band. All use $U_\infty$ as the velocity scale.}
  \label{tab:keyFrequencies}
\end{table}

\subsection{High-frequency centreline shear–layer shedding: $f\simeq 135$, $200$~Hz}

The upper frequency band corresponds to classical roll-up and advection in the separated shear layer. It is strongest for centreline taps that straddle the mean separating streamline and persists downstream along the symmetry plane (see wall-pressure PSD maps in figure~\ref{fig:CpRMS_spectra}). Non-dimensionalization of this frequency using $H$ gives $St_H\simeq 0.23$; using the separation length gives $St_{L_{\rm sep}}\simeq 0.68$, squarely within the $0.5$–$1.0$ range reported for fixed-separation shear layers \citep{eaton1982,kiya1983,cherry1984}. The side-on POD modes associated with this band exhibit canonical roll-up/advection mode shapes in both $u$ and $w$. 

\vspace{0.5em}
\noindent
Within this band we observe higher-frequency content near $200$~Hz close to separation and a lower, dominant peak near $135$~Hz farther downstream. We interpret the $\sim\!200$~Hz component as the nascent Kelvin–Helmholtz roll-up set by the local shear-layer thickness and convective speed in the separation region; as vortices convect and amalgamate, their spacing increases and the effective passage frequency downshifts toward $\sim\!135$~Hz, which represents the mature, large-scale shedding of the vortex packets. This progressive downshift is consistent with vortex amalgamation dynamics commonly observed in separated shear layers \citep{cherry1984}.

\subsection{Breathing mode: $f\simeq 13.5$~Hz}
The 13.5~Hz band is attributed to a breathing motion of the separated zone at the centreline. In the side-on PIV plane, the associated POD mode displays a modulation of the reverse-flow extent and a streamwise displacement of the separation point, i.e.\ expansion/contraction of the recirculation (``breathing'').
At this frequency the corresponding Strouhal number is $St_{L_{\rm sep}} = 0.068$, slightly below—but of the same order as—the canonical range of the flapping mode reported for fixed-separation shear layers,
$St_{L_{\rm sep}}\in[0.08,\,0.20]$ \citep{fang2024}. The value is also consistent with the breathing mode in variable-separation flows in which the separation point is free to fluctuate: it is larger than the diffuser value of \citet{weiss2015} ($St_{L_{\rm sep}}=0.01$) and close to the trailing-edge separated wing of \citet{wang2022} ($St_{L_{\rm sep}}=0.05$). Thus, the breathing mode identified for the \textit{Gaussian Bump} sits within the canonical range for turbulent separated flows. To our knowledge, this constitutes the first direct observation of a breathing mode in separated flows over a hill- or bump-type geometry.

\vspace{0.5em}
\noindent
Recent work by \citet{fang2024} argues that breathing/flapping is a universal feature of separated flows, clustering in a narrow band of $St_{L_{\rm sep}}\in[0.08,\,0.20]$ and consistent with an absolute (global) instability localized near the crest of the mean separating streamline (lift-up amplified by adverse streamline curvature). Our observation of the breathing mode at $St_{L_{\rm sep}}=0.068$ falls just below this interval but remains of the same order and exhibits the same kinematics.

\vspace{0.5em}
\noindent
Moreover, the breathing mode is most naturally associated with the symmetric family of the top-down POD modes, particularly the elongation/shortening captured by \(a_{\text{S}}^{(1)}\); consistently, the joint distributions in figure~\ref{fig:keyPoints_jointPDF} show that $a^{(1)}_{\rm S}$ is maximal and positive near $a^{(1)}_{\rm A}\!\approx\!0$ (most elongated symmetric state) and becomes negative as $\lvert a^{(1)}_{\rm A}\rvert$ grows (contracted wake/upstream state). This picture echoes the ``pumping'' mode described by \citet{pavia2019} for bluff-body wakes, where stretching/squeezing of a hairpin-like structure produces a streamwise bubble-breathing motion. Higher-rate, synchronized pressure–PIV or volumetric PIV would allow direct phase-resolved explorations.

\vspace{0.5em}
\noindent
Recent studies of TSBs have suggested that the low-frequency breathing motion observed in the symmetry plane of 2D geometries often reflects an underlying spanwise-structured stationary 3D global mode. Experiments on backward-facing ramps \citep{steinfurth2025,fuchs2025} and wall-mounted humps \citep{borgmann2024} reveal spanwise standing-wave patterns whose nodes and antinodes modulate the extent of the reverse-flow region, producing the familiar \textit{breathing} motion when viewed in a single plane. 
These recent results additionally demonstrate that the spanwise domain size (here, tunnel width $L$) emerges as a second relevant length scale alongside the mean separation length $L_{\rm{sep}}$ when interpreting the low-frequency unsteadiness.
While the present bump measurements capture sufficient spanwise coverage of the instantaneous separated zone, they do not provide the time-resolved 3D information needed to identify standing-wave organization or spanwise eigenmodes in the sense of \cite{borgmann2024,steinfurth2025,fuchs2025}. \cite{klopsch2025} recently performed resolvent analysis on an experimentally assimilated 2D mean flow of the \textit{Gaussian Bump}, imposing spanwise-periodic boundary conditions that permit harmonic 3D perturbations. Their analysis identified a low-frequency 3D stationary global mode within a similar frequency range to that shown here, further supporting this standing-wave concept.

\subsection{Very-low-frequency spanwise meandering: $f\simeq 1$~Hz\label{sec:VLF}}
The slowest VLF band appears as a broadband hump in the premultiplied spectrum of the dominant antisymmetric top–down mode \(a_{\text{A}}^{(1)}\) (figure~\ref{fig:a1_psd_pdf}a) as well as the off–centreline tap spectra near the separated region away from the centerline (figure~\ref{fig:CpRMS_spectra}a,d,f). For this frequency, $St_H\simeq 1.7\times 10^{-3}$, which is consistent with the order reported for slow lateral wake motions in other 3D separated flows. In square–back Ahmed bodies, $St_H\sim 10^{-3}$ characterizes large-scale wake switching between spanwise-asymmetric states \citep{grandemange2013}; the BeVERLI hill ($B_w/H=5$) likewise switches at the same order \citep{macgregor2023}. For axisymmetric bodies, the diameter $D$ is the appropriate scale and \citet{zhang2023} report a near-wake barycentre precession of the planar vortex-shedding axis at $St_D = 0.002$–$0.005$, consistent with the $\mathcal{O}(10^{-3})$ azimuthal meandering frequencies reported by \citet{rigas2014}.

\vspace{0.5em}
\noindent
Despite the similarity in Strouhal number across configurations, the present spanwise motion is not a bi- or multi-stable switching phenomenon. Two lines of evidence indicate a \emph{meandering} process about the symmetric mean. First, the PDF of the antisymmetric wall-pressure combination $a=\tfrac{1}{2}(p_2-p_1)$ is approximately Gaussian for all symmetric tap pairs (figure~\ref{fig:a1_psd_pdf}b). The PDF of the leading antisymmetric POD coefficient $a^{(1)}_{\rm A}$ shows only mild positive skew ($0.17\pm0.04$) and sub-Gaussian kurtosis ($2.43\pm0.08$), with no hint of the bimodality or heavy tails characteristic of long residence in multiple metastable states. Second, the joint distributions in $\{a_{\text{A}}^{(1)}, a_{\text{S}}^{(1-3)}\}$-space (figure~\ref{fig:keyPoints_jointPDF}) form a single, concentrated density centered near the mean symmetric state at $a_{\text{A}}^{(1)}\!\approx\!0$, with no evidence of two well-separated high-density lobes of the kind expected for bistable switching. This indicates that the wake occupies a neighbourhood of the mean state rather than dwelling at two distinct lateral configurations. Consistently, instantaneous reconstructions of $a_{\text{A}}^{(1)}$ superimposed on the mean field evolve smoothly in time (see \textbf{Supplementary Movie 1: A1}), again supporting a continuously varying meandering motion rather than abrupt transitions between preferred states.

\vspace{0.5em}
\noindent
The observed absence of bistable switching in the present flow could be influenced by the geometry. The present bump is wide ($B_w/H\approx 8.3$), whereas canonical bistable configurations---e.g., Ahmed bodies \citep{grandemange2013}, axisymmetric hills \citep{garcia2009}, half–axisymmetric surface–mounted bodies \citep{panesar2023}, and the BeVERLI hill \citep{duetsch2022,macgregor2023}---have unit to moderate aspect ratios that appear to promote interaction between the two lateral shear layers. 
That the same order of frequency ($St\sim 10^{-3}$) emerges across diverse configurations is nevertheless striking and suggests a common large–scale driver, as argued for bluff–body wakes by \citet{pavia2019}. In particular, the recurrence of $St\sim10^{-3}$ across these geometries indicates that VLF lateral motions arise generically in separated 3D wakes, while the specific manifestation—meandering versus switching—depends on the geometry and boundary conditions.

We note that this geometry exhibits two low-frequency motions (i.e., the spanwise meandering VLF motion and the low-frequency breathing motion), whereas many flows in which low-frequency breathing has been explored reveal only a single frequency. This is perhaps not surprising, since the present geometry has two orthogonal lengthscales (height and width), whereas most previous studies have focused on 2D configurations. 
Intriguingly, scaling the VLF frequency with two key spanwise length scales: the bump half-maximum width $B_w$ and the tunnel width $L$ (as motivated by the work of \cite{borgmann2024, steinfurth2025})—gives $St_{B_w}\approx 0.014=\mathcal{O}(10^{-2})$ and $St_{L}\approx 0.0203=\mathcal{O}(10^{-2})$, respectively. Both are of the same order as the 13.5~Hz breathing frequency when it is scaled with either $H$ or $L_{\rm sep}$ (yielding $St_H=0.023$ and $St_{L_{\rm sep}}=0.068$). 
While it might be tempting to suggest a shared mechanism, the absence of a VLF peak on the centreline pressure spectra argues against the notion that the 13.5~Hz breathing motion observed in the side–on PIV is simply a projection of the spanwise meandering onto the symmetry plane. Instead, the two modes appear to be dynamically distinct: the 1~Hz VLF mode reflects a large-scale spanwise oscillation of the separated footprint, whereas the 13.5~Hz mode corresponds to its streamwise–vertical breathing motion. We therefore regard the similarity of their Strouhal numbers as an intriguing observation and look forward to future studies that vary the bump width so that its influence on the low-frequency dynamics can be more fully characterized.

\subsection{Lateral shear-layer shedding: $f\simeq20$~Hz}
The final energetic band, centred near $20$~Hz, manifests distinctly from the VLF 1~Hz spanwise meandering, the 13.5 Hz breathing, and the 135~Hz high-frequency centreline shedding modes. In the pressure spectra (figure~\ref{fig:CpRMS_spectra}c), this band is sharply localized to the spanwise plane containing the surface-vortex cores. It appears prominently on the off-centreline taps flanking each core (taps 23/24) and with comparable spectral amplitude on the corresponding centreline tap on the same spanwise plane (tap 9) at $x/H=1.47$, suggesting a coherent process tied to that local 3D structure. The cross-correlations between taps 23 and 24 exhibit a strong positive peak at zero-lag, indicating that the motions associated with this band are in-phase across the centreline (figure~\ref{fig:xcorr}a).

\vspace{0.5em}
\noindent
The spatial localization of the 20~Hz signature, together with the known topology of the time-averaged 3D flow (figure~\ref{fig:3Drendering}a,b), points to a mechanism involving the \emph{lateral} shear layers that bound the separated footprint. These shear layers wrap around the surface-vortex cores and remain close enough to the wall near the vortex-core plane at $x/H\approx1.47$ to imprint strongly on the wall-pressure field. Their proximity on both sides of the centreline at the vortex-core plane could explain why taps 9, 23, 24 all show comparable spectral energy at 20~Hz.

\vspace{0.5em}
\noindent
Downstream of the vortex-core plane, the 20~Hz band rapidly diminishes and is absent beyond the local surface-concavity change ($x/H\!\approx\!1.65$). This attenuation is consistent with the increasing wall-normal displacement of the lateral shear layers as the flow progresses downstream: as these shear layers lift away from the wall, their wall-pressure imprint weakens and eventually falls below detectability. Taken together, the evidence supports interpreting the 20~Hz band as a localized vortex-shedding process associated with the lateral shear layers near the surface-vortex cores.

\section{Conclusions\label{sec:Conclusions}}
The unsteady dynamics of turbulent separation over the three-dimensional \textit{Boeing Gaussian Bump} were characterized at $\Rey_H=2.26\times10^5$ (with bump height $H$) using synchronized wall-pressure measurements and planar particle image velocimetry (PIV) in orthogonal side-on and top-down planes. 
The side-on PIV reveals a streamwise-elongated separation zone along the centreline with a mean separation length $L_{\rm{sep}}\approx 3H$ that spans the region approximately bounded by the upstream and downstream centreline surface saddle points. The separated region is laterally confined by two surface nodes corresponding to the footprints of wall-normal vortices that turn toward the downstream direction, and the mean flow exhibits a centreline upwash downstream of the bump. Together, these mean-flow features are consistent with the canonical owl-face-of-the-first-kind topology.

\vspace{0.5em}
\noindent
The primary focus of this study is the unsteady features that contribute to this three-dimensional structure and their dynamical characteristics. Four dominant spectral bands in the wall-pressure  were identified. At the highest frequencies (135, 200~Hz), the centreline shear layer exhibits vortex shedding behavior. Side-on PIV reveals that the upper band near 200~Hz is interpreted to be associated with the nascent Kelvin-Helmholtz instability near separation, and the progressive reduction in frequency towards 135~Hz as the flow advects downstream is thought to reflect nonlinear amalgamation of large-scale vortical structures. 
The associated scaling with mean separation length gives $St_{L_{\rm sep}}=fL_{\rm{sep}}/U_{\infty}=0.68$, which is in the range reported for the shedding mode in fixed-separation 2D turbulent separation bubbles (TSBs). Similarly, a spectral peak near $20$~Hz is found to be strongest near the surface-vortex cores and is interpreted to reflect vortex shedding from the lateral shear layers bounding this region, supported by strong phase correlation between symmetric wall-pressure taps flanking the surface nodes.

\vspace{0.5em}
\noindent
A combination of pressure signals and temporally upsampled POD modes reveals dynamics consistent with breathing-type motions observed in other separated flows. In particular, this motion is associated with a low-frequency $13.5$~Hz band, concentrated in the wall-pressure spectra for centreline taps upstream of separation. The temporally upsampled flow-fields reveal a vertical oscillation of the shear layer accompanied by a streamwise displacement of the separation point and reverse-flow extent, which are characteristics of low-frequency breathing motions. Scaling by the mean separation length yields $St_{L_{\rm sep}}\approx0.068$, which is close to the lower end of the range for flapping modes in canonical fixed-separation 2D TSBs and roughly in the range of breathing modes observed in variable-separation 2D TSBs.

\vspace{0.5em}
\noindent
Perhaps most interestingly, this study reveals the presence of a very-low-frequency (VLF) band centered near $1$~Hz ($St_H\approx1.7\times10^{-3}$) that dominates the off-centreline separated region, and is associated with an anti-symmetric motion of the separated zone. The leading antisymmetric POD mode from top-down PIV data reveals that this VLF band is associated with a continuous lateral wake oscillation without long residence in a preferred state, which we refer to as a \textit{meandering} motion. 
This contrasts with axisymmetric wakes, where meandering arises from free azimuthal precession, and with rectilinear reflectionally symmetric bodies, where bistable switching has been known to occur between two discrete spanwise-asymmetric states (depending on body aspect ratio). Critically, it is found that the meandering mode preserves a similarly slow timescale ($St_H\sim10^{-3}$) to that observed in the other 3D turbulent separated flows, despite contrasting global dynamical features.

\vspace{0.5em}
\noindent
In addition to the separation of time scales, the present data reveal a coherent organization of the inter-modal unsteady dynamics in a low-dimensional state space spanned by symmetric and antisymmetric top-down POD modes. Joint distributions between the leading antisymmetric POD coefficient (associated with the VLF spanwise meandering) and the leading symmetric coefficient show that the wake spends most of its time near the symmetric mean state, with increasingly rare excursions towards extreme asymmetric states. The low-probability rims of the distribution form an approximately parabolic envelope. Instantaneous flow-field reconstructions corresponding to time instants at the vertex of this envelope indicate that symmetric states that are rarely sampled correspond to the most streamwise-elongated symmetric wake configuration, whereas extreme antisymmetric states coincide with a shortened, laterally displaced wake that is shifted upstream relative to the mean. 
The leading symmetric mode therefore captures the dominant streamwise stretching and contraction of the wake---thus likely associated with the breathing motion---while higher-order symmetric modes appear to contribute to intermediate wake deformations that are potentially associated with the shedding of large loop vortices.

\vspace{0.5em}
\noindent
In summary, the \textit{Gaussian Bump} wake features a hierarchy of unsteady motions spanning over two decades in frequency. 
Of particular interest is the simultaneous presence of two low-frequency processes: a breathing-type modulation of the separated region and a very-low-frequency spanwise meandering of the separation zone. Despite geometric and boundary-layer differences, the frequency scaling and modal organization of these motions align closely with those in other turbulent separated flows, suggesting that there exists a common set of interacting instabilities/motions under broadly varying conditions.
\appendix
\section{}\label{appA}

\subsection{Side-on flow-field turbulence\label{sec:RD_flowfieldTurbulence}}



\begin{figure}
    \centering
    \includegraphics[width=\textwidth]{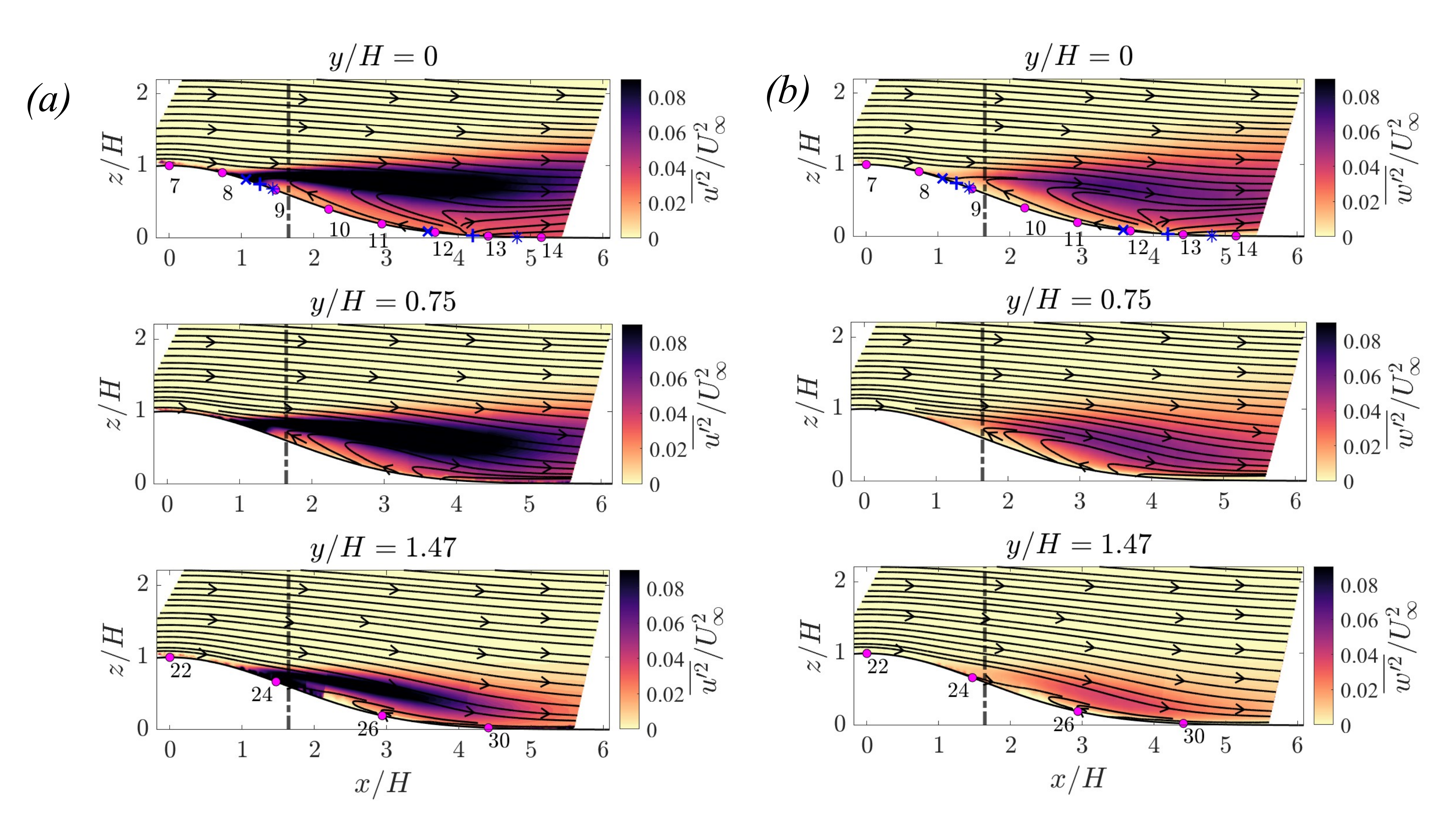}
    \caption{\small{Normal Reynolds stresses (a) $\overline{u'^2}$, (b) $\overline{w'^2}$ along the $x$-$z$ plane at three spanwise locations (from top to bottom) $y/H\,=\,0,\,0.73,\,1.47$. Vertical dotted line denotes location of surface concavity change}}
    \label{fig:yPos_stresses_norm}
\end{figure}

The normal Reynolds stresses $\overline{u'^2},\,\overline{w'^2}$ along three streamwise-vertical planes at $y/H=0,\,0.73,\,1.47$ are shown along with the mean streamlines in figure \ref{fig:yPos_stresses_norm}. At the centreline $y/H=0$, the streamwise component $\overline{u'^2}$ dominates the shear layer region in a streamwise-parallel band that emanates from the separation point at $x/H\approx1.25$, and extends downstream (figure \ref{fig:yPos_stresses_norm}a). Hence, this suggests that $\overline{u'^2}$ is first generated through the shear layer. Moving outboard, the shear layer progressively tilts down towards the wall. A maximum shear layer thickness of $\Delta\,z\approx0.5H$ is achieved in the range $x/H\in[3,4]$, and is observed both at the centreline $y/H=0$ as well as at $y/H=0.73$. At $y/H=1.47$, the shear layer starts thinning. At all spanwise planes, $\overline{u'^2}$ starts to diffuse at the location of the downstream saddle point of the bifurcation (end of the reverse-flow region); for instance, see $x/H>4$ for the $y/H=0$ case in figure \ref{fig:yPos_stresses_norm}a. At the centreline, the diffusion process is aided by the upwash observed at $x/H>4.5$.


\vspace{0.5em}
\noindent
The vertical component of the Reynolds stresses $\overline{w'^2}$ shows a similar trend, with a few notable differences (figure \ref{fig:yPos_stresses_norm}b). First, the magnitude of $\overline{w'^2}$ is nearly half that of $\overline{u'^2}$, which is expected due to the presence of the wall. Second, the high-magnitude region of $\overline{w'^2}$ begins just downstream of the location of surface concavity change at $x/H=1.65$, as opposed to $\overline{u'^2}$, which starts further upstream at the separation point. The $\overline{w'^2}$ magnitude also decreases outboard at $y/H=1.47$.

\begin{figure}
    \centering
    \includegraphics[width=\textwidth]{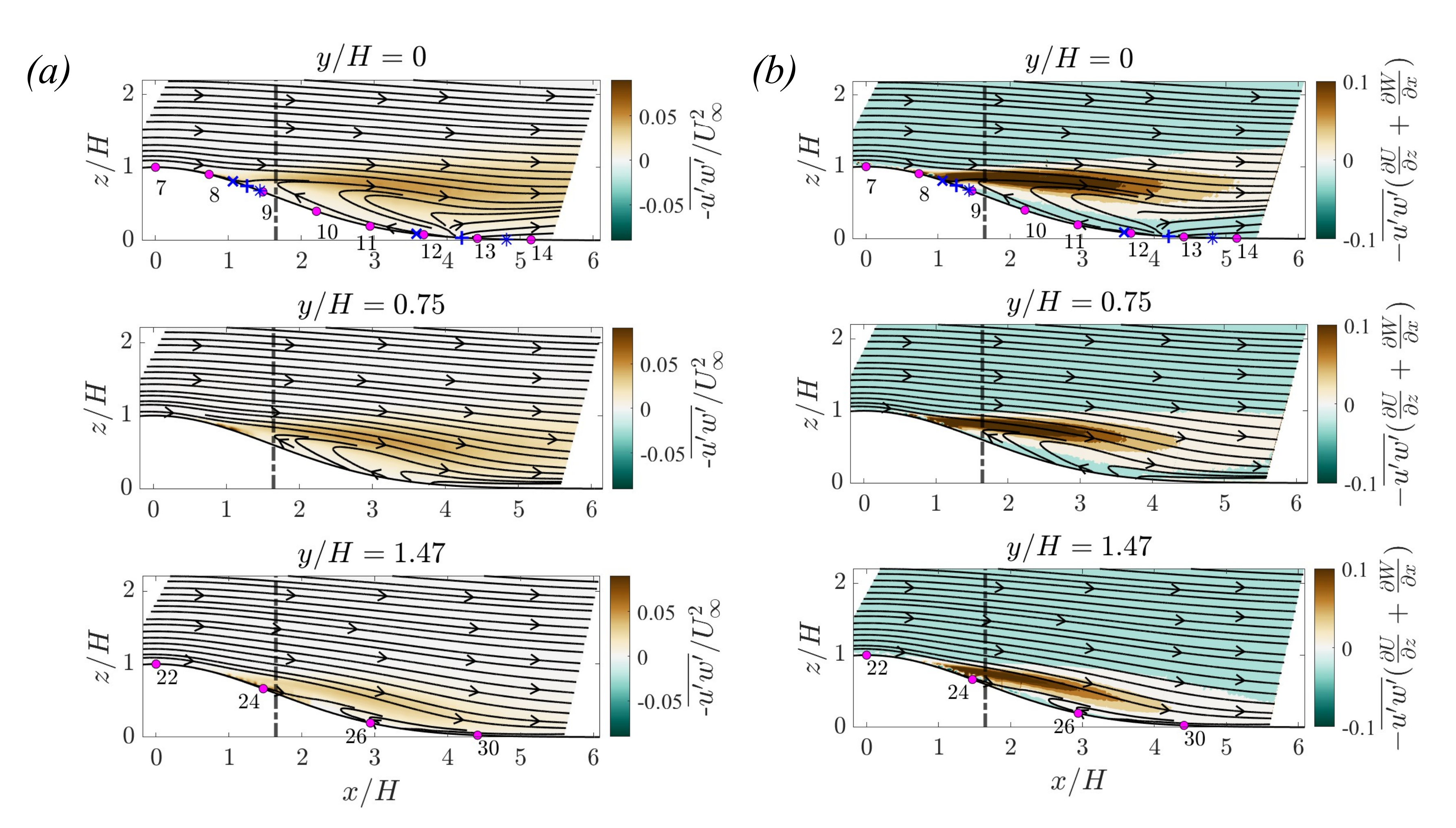}
    \caption{\small{(a) Reynolds shear stress $\overline{u'w'}$, (b) contribution of $\overline{u'w'}$ to turbulence production, denoted by $-\overline{u'w'}\,(\partial U/\partial z + \partial W / \partial x)$. Formatted similarly to figure \ref{fig:yPos_stresses_norm}}}
    \label{fig:yPos_uw_prod}
\end{figure}

\vspace{0.5em}
\noindent
Contours of the Reynolds shear stress $\overline{u'w'}$ show that it is most strongly expressed around the shear layer region for $x/H\in[2, 4]$ (figure \ref{fig:yPos_uw_prod}a). Moreover, just like the normal stresses, $\overline{u'w'}$ starts to diffuse when the flow approaches the downstream saddle point of the bifurcation, aided by centreline upwash for $x/H>4.5$.


\vspace{0.5em}
\noindent
The Reynolds shear stress $\overline{u'w'}$ is generated through different mechanisms compared to the normal stresses $\overline{u'^2},\,\overline{w'^2}$. While the streamwise component $\overline{u'^2}$ is likely generated through the shear layer, the action of the shear component $\overline{u'w'}$ is to generate the turbulence within it through the mean velocity gradients. The contribution of $\overline{u'w'}$ to the turbulence production, quantified by $-\overline{u'w'}\,(\partial U/\partial z + \partial W / \partial x)$, is provided in figure \ref{fig:yPos_uw_prod}b. As anticipated, there is positive production through $\overline{u'w'}$ along the shear layer. In particular, this turbulence-generating region is an approximately streamwise-parallel band at the centreline and progressively tilts wall-ward off-centreline.

\paragraph{
\textbf{Acknowledgments. }
KHM would like to gratefully acknowledge funding from the Natural Sciences and Engineering Research Council of Canada (NSERC) and Alberta Innovates. 
The support of Boeing Commercial Aircraft and their Flight Sciences Function Excellence group is gratefully acknowledged, especially that of Jeffrey Slotnick and Philippe Spalart for their significant technical and administrative support for this effort. The manufacture of the bump test article was conducted by Steven Seim of CyberModelle, to whom we owe a debt
of gratitude for his tireless effort.
}

\paragraph{
\textbf{Declaration of Interests. }
The authors report no conflict of interest.
}





\bibliographystyle{jfm}
\bibliography{jfm-instructions}

@article{cura2025,
  title={Linear modeling of a family of turbulent separation bubbles},
  author={Cura, Carolina and Hanifi, Ardeshir and Cavalieri, Andr{\`e} VG and Weiss, Julien},
  journal={Physical Review Fluids},
  volume={10},
  number={11},
  pages={114607},
  year={2025},
  publisher={APS}
}

@article{cura2024,
  title={On the low-frequency dynamics of turbulent separation bubbles},
  author={Cura, C and Hanifi, Ardeshir and Cavalieri, AVG and Weiss, J},
  journal={Journal of Fluid Mechanics},
  volume={991},
  pages={A11},
  year={2024},
  publisher={Cambridge University Press}
}

@article{rigas2014,
  title={Low-dimensional dynamics of a turbulent axisymmetric wake},
  author={Rigas, G and Oxlade, AR and Morgans, AS and Morrison, JF},
  journal={Journal of Fluid Mechanics},
  volume={755},
  pages={R5},
  year={2014},
  publisher={Cambridge University Press}
}

@techreport{gray2023onr,
  author      = {Patrick Gray and Thomas Corke and Flint Thomas and Igal Gluzman and Joseph Straccia},
  title       = {Turbulence Model Validation through Joint Experimental/Computational Studies of Separated Flow over a Three-Dimensional Tapered Bump: {Part I}---Experimental Investigation (Final Report)},
  institution = {University of Notre Dame},
  address     = {Notre Dame, IN, USA},
  year        = {2023},
  month       = jul,
  type        = {Final report},
  note        = {ONR Contract N00014-20-2-1002; NASA Contract 80LARC21T0001},
  url         = {https://turbmodels.larc.nasa.gov/Other_exp_Data/speedbump_sep_exp.html}
}

@article{pavia2019,
  title={Three dimensional structure of the unsteady wake of an axisymmetric body},
  author={Pavia, Giancarlo and Varney, Max and Passmore, Martin and Almond, Mathew},
  journal={Physics of Fluids},
  volume={31},
  number={2},
  year={2019},
  publisher={AIP Publishing}
}

@article{panesar2023,
  title={Oblique and parallel modes of the bistable bluff body wake},
  author={Panesar, Simran Singh and Xia, Hao and Passmore, Martin},
  journal={Physical Review Fluids},
  volume={8},
  number={8},
  pages={084601},
  year={2023},
  publisher={APS}
}

@article{perryHornung1984,
  title={Some aspects of three-dimensional separation. II-Vortex skeletons},
  author={Perry, Authony E and Hornung, Hans},
  journal={Zeitschrift fur Flugwissenschaften und weltraumforschung},
  volume={8},
  pages={155--160},
  year={1984}
}

@article{simpson1989,
  title={Turbulent boundary-layer separation},
  author={Simpson, Roger L},
  journal={Annual Review of Fluid Mechanics},
  volume={21},
  number={1},
  pages={205--232},
  year={1989},
  publisher={Annual Reviews 4139 El Camino Way, PO Box 10139, Palo Alto, CA 94303-0139, USA}
}

@article{mason1987,
  title={Trailing vortices in the wakes of surface-mounted obstacles},
  author={Mason, PJ and Morton, BR},
  journal={Journal of Fluid Mechanics},
  volume={175},
  pages={247--293},
  year={1987},
  publisher={Cambridge University Press}
}

@article{byun2006,
  title={Structure of three-dimensional separated flow on an axisymmetric bump},
  author={Byun, Gwibo and Simpson, Roger L},
  journal={AIAA Journal},
  volume={44},
  number={5},
  pages={999--1008},
  year={2006}
}

@article{manohar2023,
  title={Temporal super-resolution using smart sensors for turbulent separated flows},
  author={Manohar, Kevin H and Williams, Owen and Martinuzzi, Robert J and Morton, Chris},
  journal={Experiments in Fluids},
  volume={64},
  number={5},
  pages={101},
  year={2023},
  publisher={Springer}
}

@article{sears1975,
  title={Boundary-layer separation in unsteady flow},
  author={Sears, WR and Telionis, DP},
  journal={SIAM Journal on Applied Mathematics},
  volume={28},
  number={1},
  pages={215--235},
  year={1975},
  publisher={SIAM}
}

@inproceedings{gray2021,
  title={A new validation experiment for smooth-body separation},
  author={Gray, Patrick D and Gluzman, Igal and Thomas, Flint and Corke, Thomas and Lakebrink, Matthew and Mejia, Kevin},
  booktitle={AIAA Aviation 2021 Forum},
  pages={2810},
  year={2021}
}

@inproceedings{gray2023,
  title={Experimental and Computational Evaluation of Smooth-Body Separated Flow over Boeing Bump},
  author={Gray, Patrick D and Lakebrink, Matthew T and Thomas, Flint O and Corke, Thomas C and Gluzman, Igal and Straccia, Joseph},
  booktitle={AIAA AVIATION 2023 Forum},
  pages={3981},
  year={2023}
}

@inproceedings{williams2022,
  title={Comparison of hill-type geometries for the validation and advancement of turbulence models},
  author={Williams, Owen J and Annamalai, Hariprasad and Ozoroski, Thomas A and Roy, Christopher J and Lowe, Todd},
  booktitle={AIAA SCITECH 2022 Forum},
  pages={1032},
  year={2022}
}

@inproceedings{bradshaw1973,
  title={Effects of streamline curvature on turbulent flow},
  author={Bradshaw, Peter and Young, AD},
  year={1973},
  organization={Agard Paris}
}

@article{kiya1983,
  title={Structure of a turbulent separation bubble},
  author={Kiya, Masaru and Sasaki, Kyuro},
  journal={Journal of Fluid Mechanics},
  volume={137},
  pages={83--113},
  year={1983},
  publisher={Cambridge University Press}
}

@inproceedings{eaton1982,
  title={Low frequency unsteadyness of a reattaching turbulent shear layer},
  author={Eaton, John K and Johnston, James P},
  booktitle={Turbulent Shear Flows 3: Selected Papers from the Third International Symposium on Turbulent Shear Flows, The University of California, Davis, September 9--11, 1981},
  pages={162--170},
  year={1982},
  organization={Springer}
}

@article{na1998,
  title={The structure of wall-pressure fluctuations in turbulent boundary layers with adverse pressure gradient and separation},
  author={Na, Y and Moin, Parviz},
  journal={Journal of Fluid Mechanics},
  volume={377},
  pages={347--373},
  year={1998},
  publisher={Cambridge University Press}
}

@article{na1998dns,
  title={Direct numerical simulation of a separated turbulent boundary layer},
  author={Na, Y and Moin, Parviz},
  journal={Journal of Fluid Mechanics},
  volume={374},
  pages={379--405},
  year={1998},
  publisher={Cambridge University Press}
}

@article{driver1987,
  title={Time-dependent behavior of a reattaching shear layer},
  author={Driver, David M and Seegmiller, H Lee and Marvin, Joe G},
  journal={AIAA Journal},
  volume={25},
  number={7},
  pages={914--919},
  year={1987}
}

@article{spazzini2001,
  title={Unsteady behavior of back-facing step flow},
  author={Spazzini, PG and Iuso, Gaetano and Onorato, Michele and Zurlo, Nicola and Di Cicca, Gaetano Maria},
  journal={Experiments in Fluids},
  volume={30},
  number={5},
  pages={551--561},
  year={2001},
  publisher={Springer}
}

@article{weiss2015,
  title={Unsteady behavior of a pressure-induced turbulent separation bubble},
  author={Weiss, Julien and Mohammed-Taifour, Abdelouahab and Schwaab, Quentin},
  journal={AIAA Journal},
  volume={53},
  number={9},
  pages={2634--2645},
  year={2015},
  publisher={American Institute of Aeronautics and Astronautics}
}

@article{wang2022,
  title={Unsteady motions in the turbulent separation bubble of a two-dimensional wing},
  author={Wang, Sen and Ghaemi, Sina},
  journal={Journal of Fluid Mechanics},
  volume={948},
  pages={A3},
  year={2022},
  publisher={Cambridge University Press}
}

@article{bourgeois2013,
  title={Generalized phase average with applications to sensor-based flow estimation of the wall-mounted square cylinder wake},
  author={Bourgeois, JA and Noack, BR and Martinuzzi, RJ},
  journal={Journal of Fluid Mechanics},
  volume={736},
  pages={316--350},
  year={2013},
  publisher={Cambridge University Press}
}

@article{ching2020,
  title={Large-eddy simulation study of unsteady wake dynamics and geometric sensitivity on a skewed bump},
  author={Ching, David S and Eaton, John K},
  journal={Journal of Fluid Mechanics},
  volume={885},
  pages={A22},
  year={2020},
  publisher={Cambridge University Press}
}

@book{adrian2011,
  title={Particle image velocimetry},
  author={Adrian, Ronald J and Westerweel, Jerry},
  number={30},
  year={2011},
  publisher={Cambridge University Press}
}

@article{discetti2013,
  title={{PIV} measurements of anisotropy and inhomogeneity in decaying fractal generated turbulence},
  author={Discetti, Stefano and Ziskin, Isaac B and Astarita, Tommaso and Adrian, Ronald J and Prestridge, Kathy P},
  journal={Fluid Dynamics Research},
  volume={45},
  number={6},
  pages={061401},
  year={2013},
  publisher={IOP Publishing}
}

@article{klopsch2025,
  title={Spectral analysis of attached and separated turbulent flows over a Gaussian-shaped bump},
  author={Klopsch, Roman and Fuchs, Lukas M and Rigas, Georgios and Oberleithner, Kilian and von Saldern, Jakob GR},
  journal={arXiv preprint arXiv:2512.13582},
  year={2025}
}

@article{simmons2024,
  title={Experimental characterization and similarity scaling of smooth-body flow separation and reattachment on a two-dimensional ramp geometry},
  author={Simmons, DJ and Thomas, FO and Corke, TC and Gluzman, I},
  journal={Journal of Fluid Mechanics},
  volume={1000},
  pages={A12},
  year={2024},
  publisher={Cambridge University Press}
}

@article{simmons2022,
  title={Experimental characterization of smooth body flow separation topography and topology on a two-dimensional geometry of finite span},
  author={Simmons, DJ and Thomas, FO and Corke, TC and Hussain, F},
  journal={Journal of Fluid Mechanics},
  volume={944},
  pages={A42},
  year={2022},
  publisher={Cambridge University Press}
}

@article{borgmann2024,
  title={Three-Dimensional Nature of Low-Frequency Unsteadiness in a Turbulent Separation Bubble},
  author={Borgmann, David and Cura, Carolina and Weiss, Julien and Little, Jesse},
  journal={AIAA Journal},
  volume={62},
  number={11},
  pages={4349--4363},
  year={2024},
  publisher={American Institute of Aeronautics and Astronautics}
}

@article{fuchs2025,
    title={Standing-wave dynamics in low-frequency breathing of a turbulent separation bubble},
    volume={1030},
    DOI={10.1017/jfm.2026.11191},
    journal={Journal of Fluid Mechanics},
    author={Fuchs, Lukas M. and Steinfurth, Ben and von Saldern, Jakob G.R. and Weiss, Julien and Oberleithner, Kilian},
    year={2026},
    pages={A34}
}

@article{steinfurth2025,
  title={Three-dimensional low-frequency dynamics of a turbulent separation bubble},
  author={Steinfurth, Ben and Li, Mogeng and Scarano, Fulvio and Weiss, Julien},
  journal={Journal of Fluid Mechanics},
  volume={1019},
  pages={R4},
  year={2025},
  publisher={Cambridge University Press}
}

@article{williamson1996,
  title={Three-dimensional wake transition},
  author={Williamson, CHK},
  journal={Journal of Fluid Mechanics},
  volume={328},
  pages={345--407},
  year={1996},
  publisher={Cambridge University Press}
}

@article{grandemange2012a,
  title={Reflectional symmetry breaking of the separated flow over three-dimensional bluff bodies},
  author={Grandemange, Mathieu and Cadot, Olivier and Gohlke, Marc},
  journal={Physical review E},
  volume={86},
  number={3},
  pages={035302},
  year={2012},
  publisher={APS}
}

@article{grandemange2013,
  title={Turbulent wake past a three-dimensional blunt body. {Part} 1. Global modes and bi-stability},
  author={Grandemange, Mathieu and Gohlke, Marc and Cadot, Olivier},
  journal={Journal of Fluid Mechanics},
  volume={722},
  pages={51--84},
  year={2013},
  publisher={Cambridge University Press}
}

@article{zhang2023,
  title={Coherent motions in a turbulent wake of an axisymmetric bluff body},
  author={Zhang, Fengrui and Peet, Yulia T},
  journal={Journal of Fluid Mechanics},
  volume={962},
  pages={A19},
  year={2023},
  publisher={Cambridge University Press}
}

@article{garcia2009,
  title={Large-eddy simulation of separated flow over a three-dimensional axisymmetric hill},
  author={Garcia-Villalba, M and Li, N and Rodi, W and Leschziner, MA},
  journal={Journal of Fluid Mechanics},
  volume={627},
  pages={55--96},
  year={2009},
  publisher={Cambridge University Press}
}

@inproceedings{gargiulo2021,
  title={Flow field features of the {BeVERLI Hill} model},
  author={Gargiulo, Aldo and Duetsch-Patel, Julie E and Ozoroski, Thomas A and Beardsley, Colton and Vishwanathan, Vidya and Fritsch, Danny and Borgoltz, Aurelien and Devenport, William J and Roy, Christopher J and Lowe, Todd},
  booktitle={AIAA Scitech 2021 Forum},
  pages={1741},
  year={2021}
}

@inproceedings{duetsch2022,
  title={The {BeVERLI Hill} three-dimensional separating flow case: cross-facility comparisons of validation experiment results},
  author={Duetsch-Patel, Julie E and MacGregor, Daniel and Jenssen, Yngve L and Henry, Pierre-Yves and Muthanna, Chittiappa and Savio, Luca and Lavoie, Philippe and Gargiulo, Aldo and Sundarraj, Vignesh and Ozoroski, Thomas A and others},
  booktitle={AIAA SciTech 2022 Forum},
  pages={0698},
  year={2022}
}

@inproceedings{macgregor2023,
  title={Mean and Unsteady Surface-Pressure Measurements on the {BeVERLI Hill}},
  author={MacGregor, Daniel A and Gargiulo, Aldo and Duetsch-Patel, Julie E and Lavoie, Philippe and Lowe, Todd},
  booktitle={AIAA SCITECH 2023 Forum},
  pages={0468},
  year={2023}
}

@book{holmes2012,
    place={Cambridge},
    edition={2},
    title={Turbulence, Coherent Structures, Dynamical Systems and Symmetry},
    DOI={10.1017/CBO9780511919701},
    publisher={Cambridge University Press},
    author={Holmes, Philip and Lumley, John L. and Berkooz, Gahl and Rowley, Clarence W.},
    year={2012}
}

@article{plumejeau2019,
  title={Ultra-local model-based control of the square-back Ahmed body wake flow},
  author={Plumejeau, Baptiste and Delprat, S{\'e}bastien and Keirsbulck, Laurent and Lippert, Marc and Abassi, Wafik},
  journal={Physics of Fluids},
  volume={31},
  number={8},
  year={2019},
  publisher={AIP Publishing}
}

@article{pavia2018,
  title={Evolution of the bi-stable wake of a square-back automotive shape},
  author={Pavia, Giancarlo and Passmore, Martin and Sardu, Costantino},
  journal={Experiments in Fluids},
  volume={59},
  number={1},
  pages={20},
  year={2018},
  publisher={Springer}
}

@article{fabre2008,
  title={Bifurcations and symmetry breaking in the wake of axisymmetric bodies},
  author={Fabre, David and Auguste, Franck and Magnaudet, Jacques},
  journal={Physics of Fluids},
  volume={20},
  number={5},
  year={2008},
  publisher={AIP Publishing}
}

@article{fang2024,
  title={On the low-frequency flapping motion in flow separation},
  author={Fang, Xingjun and Wang, Zhan},
  journal={Journal of Fluid Mechanics},
  volume={984},
  pages={A76},
  year={2024},
  publisher={Cambridge University Press}
}

@inproceedings{robbins2021,
  title="Overview of validation completeness for {G}aussian speed-bump separated flow experiments",
  author="Robbins, Matthew L and Samuell, Madeline and Annamalai, Hariprasad and Williams, Owen J",
  booktitle="AIAA Scitech 2021 Forum",
  pages="0969",
  year="2021"
}

@inproceedings{williams2020,
  title={Experimental study of a {CFD} validation test case for turbulent separated flows},
  author={Williams, Owen and Samuell, Madeline and Sarwas, E Sage and Robbins, Matthew and Ferrante, Antonino},
  booktitle={AIAA Scitech 2020 Forum},
  pages={0092},
  year={2020}
}

@inproceedings{williams2021,
  title={Characterization of separated flowfield over {G}aussian speed-bump {CFD} validation geometry},
  author={Williams, Owen J and Samuell, Madeline and Robbins, Matthew L and Annamalai, Hariprasad and Ferrante, Antonino},
  booktitle={AIAA Scitech 2021 Forum},
  pages={1671},
  year={2021}
}

@Mastersthesis{annamalai2022,
  title={{Detailed characterization of turbulent separated flow dynamics and boundary layer evolution over a speed-bump geometry}},
  author={Annamalai, Hariprasad},
  school={University of Washington},
  year={2022}
}

@article{taifour2016,
  title={Unsteadiness in a large turbulent separation bubble},
  author={Mohammed-Taifour, Abdelouahab and Weiss, Julien},
  journal={Journal of Fluid Mechanics},
  volume={799},
  pages={383--412},
  year={2016},
  publisher={Cambridge University Press}
}

@article{wu2020,
  title={Spatio-temporal dynamics of turbulent separation bubbles},
  author={Wu, Wen and Meneveau, Charles and Mittal, Rajat},
  journal={Journal of Fluid Mechanics},
  volume={883},
  year={2020},
  publisher={Cambridge University Press}
}

@techreport{vision2030,
  title={{CFD} vision 2030 study: a path to revolutionary computational aerosciences},
  author={Slotnick, Jeffrey P and Khodadoust, Abdollah and Alonso, Juan and Darmofal, David and Gropp, William and Lurie, Elizabeth and Mavriplis, Dimitri J},
  institution={Natl. Aeronaut. Space Admin.},
  type={Tech. Rep.},
  number={CR-2014-218178},
  year={2014},
  address={Langley Research Center, Hampton, Virginia},
}

@inproceedings{slotnick2019,
  title={Integrated {CFD} validation experiments for prediction of turbulent separated flows for subsonic transport aircraft},
  author={Slotnick, Jeffrey P},
  booktitle={NATO Science and Technology Organization, Meeting Proceedings RDP, STO-MP-AVT-307},
  year={2019}
}

@InCollection{tropea,
  author      = {Nobach, Holger and Tropea, Cameron and Cordier, Laurent and Bonnet, Jean-Paul and Delville, Joël and Lewalle, Jacques and Farge, Marie and Schneider, Kai and Adrian, Ronald J. },
  title       = {Review of Some Fundamentals of Data Processing},
  editor      = {Tropea, Cameron and Yarin, Alexander L and Foss, John F},
  booktitle   = {Handbook of experimental fluid mechanics},
  publisher   = {Springer},
  address     = {New York},
  year        = {2007},
  pages       = {1346--1348},
  chapter     = {22},
}

@article{hudy2003,
  title={Wall-pressure-array measurements beneath a separating/reattaching flow region},
  author={Hudy, Laura M and Naguib, Ahmed M and Humphreys Jr, William M},
  journal={Physics of Fluids},
  volume={15},
  number={3},
  pages={706--717},
  year={2003},
  publisher={American Institute of Physics}
}

@inproceedings{gray2022,
  title={Experimental Characterization of Smooth Body Flow Separation Over Wall-Mounted {Gaussian} Bump},
  author={Gray, Patrick D and Gluzman, Igal and Thomas, Flint O and Corke, Thomas C},
  booktitle={AIAA SCITECH 2022 Forum},
  pages={1209},
  year={2022}
}

@article{westerweel2005,
  title={Universal outlier detection for {PIV} data},
  author={Westerweel, Jerry and Scarano, Fulvio},
  journal={Experiments in Fluids},
  volume={39},
  number={6},
  pages={1096--1100},
  year={2005},
  publisher={Springer}
}

@article{cherry1984,
  title={Unsteady measurements in a separated and reattaching flow},
  author={Cherry, NJ and Hillier, R and Latour, MEMP},
  journal={Journal of Fluid Mechanics},
  volume={144},
  pages={13--46},
  year={1984},
  publisher={Cambridge University Press}
}

@phdthesis{williams2014,
  title={Density effects on turbulent boundary layer structure: from the atmosphere to hypersonic flow},
  author={Williams, Owen JH},
  year={2014},
  school={Princeton University}
}

\end{document}